\documentclass[%
 reprint,
superscriptaddress,
amsmath,
amssymb,
aps,
pra,
longbibliography
]{revtex4-2}

\usepackage[version=3]{mhchem} 
\usepackage{lipsum}

\usepackage{longtable}

\usepackage{placeins}
\usepackage{xspace}
\usepackage{appendix}
\usepackage[detect-all]{siunitx}
\usepackage[normalem]{ulem}
\usepackage{multirow}
\usepackage{booktabs}
\usepackage{colortbl}

\usepackage{graphicx}
\usepackage{dcolumn}
\usepackage{bm}
\usepackage{booktabs}
\usepackage{hyperref}
\usepackage{xcolor}
\hypersetup{
    colorlinks,
    linkcolor={blue!70!black},
    citecolor={blue!70!black},
    urlcolor={blue!70!black}
}
\usepackage{dcolumn}
\newcolumntype{L}{D{.}{.}{2,5}}

\newcommand*{\wn}{\ensuremath{\text{cm}^{-1}}\xspace}

\newcommand*{\Eeff}{\ensuremath{\mathcal{E}_\text{eff}}\xspace}

\newcommand*{\XPi}{\ensuremath{X\,^2\Pi}\xspace}
\newcommand*{\XPiHalf}{\ensuremath{X\,^2\Pi_{1/2}}\xspace}
\newcommand*{\XPiThreeHalf}{\ensuremath{X\,^2\Pi_{3/2}}\xspace}

\newcommand*{\ASigma}{\ensuremath{A\,^2\Sigma^+}\xspace}

\begin{document}

\title{Optical Cycling and Sensitivity to the Electron's Electric Dipole Moment in Gold-Containing Molecules}
\author{K. Cooper Stuntz}
\affiliation{Department of Chemistry, Williams College, Williamstown, MA 01267, USA}
\author{Kendall L. Rice}
\affiliation{Department of Chemistry, Williams College, Williamstown, MA 01267, USA}
\author{Lan Cheng}
\affiliation{Department of Chemistry, The Johns Hopkins University, Baltimore, MD 21205, USA}
\author{Benjamin L. Augenbraun}
\email{bla1@williams.edu}
\affiliation{Department of Chemistry, Williams College, Williamstown, MA 01267, USA}

\date{July 27, 2024}

\begin{abstract}
\noindent We propose diatomic molecules built from gold and carbon-group atoms as promising candidates for optical cycling and precision measurements. We show that this class of molecules (AuX, X = C, Si, Ge, Sn, Pb) features laser-accessible electronic transitions with nearly diagonal Franck-Condon factors. The $^2\Pi_{1/2}$ ground states can be easily polarized in the laboratory frame and have near-zero magnetic moments, valuable features for quantum science and precision measurement applications. The sensitivities of AuX molecules to the electron electric dipole moment (EDM) are found to be favorable, with effective electric fields of 10--30~GV/cm. Together, these features imply that AuX molecules may enable significantly improved searches for time-reversal symmetry violation. 
\end{abstract}

\maketitle

\section{Introduction}

High-precision molecular spectroscopy is a powerful method to search for breakdowns in the fundamental symmetries of nature~\cite{demille_diatomic_2015, safronova2018search}. A leading example is the search for electric dipole moments of fundamental particles, like the electron, which has profound implications in constraining physical theories beyond the Standard Model~\cite{commins2007electric, demille_quantum_2024}. The current limit on the electron's electric dipole moment (EDM), set by searching for minute energy shifts in the spectrum of HfF$^+$, places stringent limits on the possibility of T-violating interactions at the TeV scale~\cite{roussy_improved_2023}. The tremendous energy reach of this and other~\cite{acme_collaboration_improved_2018} electron EDM searches relies on the ease with which some polar molecules can be fully polarized to give access to their large effective electric fields ($\Eeff>$10~GV/cm) in the laboratory frame. 

Molecular spectroscopy and electronic structure theory play a critical role in the identification of molecules most likely to enable new, and increasingly precise, electron EDM searches. This means identifying molecules that enjoy certain properties that enhance the sensitivity and robustness of future experiments. One such feature is a large intrinsic sensitivity to the electron EDM, characterized by a so-called effective electric field ($\Eeff$), which can be enhanced in molecular states that have large spin density on a high-$Z$ nucleus~\cite{safronova2018search}. Because EDM searches aim to measure an energy shift that is proportional to \Eeff, this preferences the use of heavy radicals. A second valuable feature, which has been used in recent limit-setting electron EDM experiments, is the presence of ground-state $\Lambda$-doublets (or, more generically, $\Omega$-doublets)~\cite{meyer_candidate_2006}. These parity doublets allow the molecules to be fully polarized and enable ``internal comagnetometry,'' in which the EDM interaction can be reversed without altering any laboratory fields. The resulting control over the sign of the EDM energy shifts can make experiments very robust against common systematic errors~\cite{eckel_search_2013, hutzler_polyatomic_2020}. Ideal parity-doubled molecular states include $^2\Pi_{1/2}$ or $^3\Delta_1$ states because these levels exhibit near-zero magnetic moments, limiting other sources of systematic error. The impressive sensitivity of electron EDM experiments using ThO~\cite{the_acme_collaboration_order_2014, acme_collaboration_improved_2018} and HfF$^+$~\cite{roussy_improved_2023, cairncross_precision_2017} have made use of these two features in low-lying $^3\Delta_1$ states.

Recent progress in molecular cooling and trapping introduces a third desirable feature: optical cycling, the repeated absorption and emission of photons among a small number of quantum states~\cite{mccarron_laser_2018, fitch2021lasercooleda, fitch_methods_2020}. Optical cycling enables unit-efficiency state preparation and readout, and it opens the door to laser cooling and trapping. The orders-of-magnitude increase in spin coherence times possible in laser cooled samples would lead to a dramatic increase in the sensitivity of future electron EDM experiments. Molecules amenable to optical cycling possess diagonal Franck-Condon factors, which ensure that optical excitation and emission can occur without populating a large number of vibrational states that are inaccessible (``dark'') to the laser light. The diatomic molecules YbF~\cite{smallman_radiative_2014}, BaF~\cite{chen_structure_2016}, and RaF~\cite{isaev_laser-cooled_2010} all exhibit good electron EDM sensitivity and the capacity for optical cycling. However, the electronic structure of these molecules precludes the existence of ground-state $\Lambda$-doublets. By contrast, molecules like ThO, ThF$^+$~\cite{ng_spectroscopy_2022}, and HfF$^+$ do possess long-lived parity-doubled states, but lack the electronic structure that enables optical cycling. Unfortunately, it has thus far proven tremendously difficult to identify molecules that possess large $\Eeff$, ground-state $\Lambda$-doublets, and optical cycling transitions~\cite{kozyryev2017precision}.

In this manuscript, we identify a class of diatomic molecules that simultaneously possesses all of these features. We propose the class of diatomic molecules AuX (X = C, Si, Ge, Sn, and Pb) as especially promising for next-generation tests of fundamental symmetry violation.  We have performed high-level electronic structure calculations of the lowest several electronic states of AuX. The potential energy curves are used to compute vibrational wavefunctions from which we extract vibrational branching fractions (VBFs) relevant to optical cycling. We also compute the effective electric fields (\Eeff), providing  insight into the suitability of each molecule for measurements of fundamental symmetry violation. Each of these molecules is predicted to possess $\Eeff$ on the order of 10~GV/cm, vibrational branching fractions that allow optically cycling, and a parity-doubled ground state with near-zero magnetic moment. This combination of features has, to the best of our knowledge, not been previously identified in other diatomic molecules.

\section{Methods}

Our computations focus on the energies and properties of the \XPi and \ASigma states of the AuX molecules. All calculations were conducted using the CFOUR suite of programs~\cite{CFOUR,Matthews2020,Stanton91a,Liu18b}.
Unless otherwise stated, we have used the exact two-component theory \cite{Dyall97,Kutzelnigg05,Liu09} with atomic mean field \cite{Hess96a} integrals (the X2CAMF scheme) \cite{liu_atomic_2018,zhang_atomic_2022, zhang_new_2024} to treat relativistic effects throughout the calculations. The X2CAMF scheme provides variational treatments of spin-orbit coupling; the computations adopt the spinor representation with spin-orbit coupling included in the orbitals. The calculations of potential energy surfaces have used Dyall's correlation-consistent triple-zeta (cc-pVTZ-SO) basis sets for the heavy elements including Au~\cite{Dyall04}, Pb, and Sn~\cite{Dyall06} recontracted for the X2CAMF scheme. These X2CAMF basis sets feature separate contraction coefficients for the spin-orbit components \cite{zhang_new_2024} and thus can account for spin-orbit coupling effects in heavy elements accurately. The correlation-consistent basis sets of C and Si \cite{Dunning89,Peterson02} recontracted for the spin-free X2C theory in its one-electron variant (the SFX2C-1e scheme) \cite{Dyall01,Liu09} have been used, since spin-orbit coupling does not affect the wave functions of these light elements much. 
The calculations of effective electric fields have used uncontracted ANO-RCC basis sets \cite{Faegri01,Roos04,Roos05} to ensure sufficient flexibility in the core region of the heavy elements and have exploited the recent implementation of analytic X2CAMF coupled-cluster (CC) and equation-of-motion coupled-cluster (EOM-CC) gradients~\cite{liu21,zhang_calculations_2021,Zhang23}.

\begin{figure}[tb]
    \centering 
    \includegraphics[width=0.95\columnwidth]{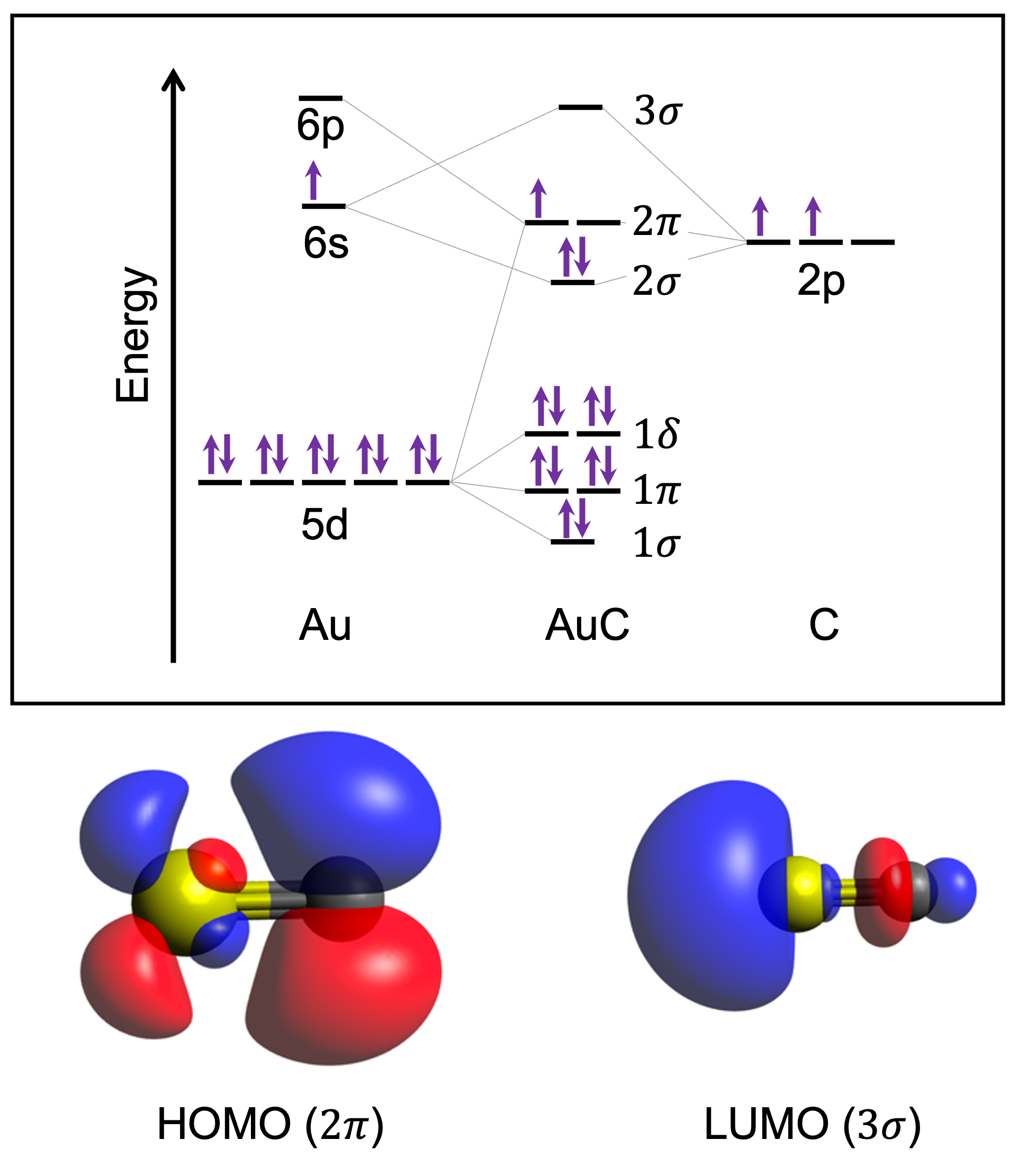}
    \caption{Molecular orbital correlation diagram and frontier molecular orbitals for the AuC radical ground state. The highest occupied ($2\pi$) and lowest unoccupied ($3\sigma$) molecular orbitals (HOMO and LUMO, respectively) are plotted at isosurface values of 0.04 a.u. The energy level diagrams and MOs of the heavier congeners are qualitatively similar.}
    \label{fig:AuC_MOs}
\end{figure}

In these molecules, the electronic configurations of the \XPiHalf, \XPiThreeHalf, and \ASigma states originate from the same closed-shell cationic configuration, differing in whether the radical electron is placed in the $2\pi_{1/2}$, $2\pi_{3/2}$, or $3\sigma$ valence spinor, respectively (see Fig.~\ref{fig:AuC_MOs}). We have therefore used equation-of-motion electron attachment coupled-cluster singles and doubles (EOMEA-CCSD) method \cite{Stanton93a,Nooijen95} to provide a balanced description of these three low-lying electronic states. The EOMEA-CCSD calculations are straightforward to converge due to stability of the cationic reference. Our calculations have used the continuum-orbital trick \cite{Stanton99} together with the available X2CAMF-EOM-CCSD program \cite{Asthana19} for excitation energies. 
The EOMEA-CCSD method has previously been shown to satisfactorily model the optical cycling properties for YO~\cite{zhang_towards_2020} and alkaline-earth monohydroxides~\cite{zhang_accurate_2021,mengesha2020branching}.

The \XPiHalf and \XPiThreeHalf states are the lowest electronic states with $\Omega=1/2$ and $\Omega=3/2$, respectively. They can be optimized directly in the UHF-based calculations. We have therefore also performed Kramers unrestricted Hartree-Fock (UHF) based CCSD \cite{Purvis82} and CCSD augmented with non-iterative triples [CCSD(T)] \cite{Raghavachari89} calculations for the \XPiHalf and \XPiThreeHalf states of the AuX molecules. The UHF-CCSD and CCSD(T) calculations directly optimize the molecular spinors for the targeted states and are likely to provide more accurate predictions of energies and properties than the EOMEA-CCSD calculations. 
To locate the energy of an expected low-lying $a\,^4\Sigma^-$ state, we have also carried out a series of SFX2C-1e restricted open-shell HF (ROHF) CCSD and CCSD(T) calculations using cc-pVTZ basis sets with scalar-relativistic contraction \cite{Peterson2005}. These calculations, which neglected spin-orbit coupling, were performed by converging to the lowest energy quartet state. Comparison to SFX2C-1e-ROHF-CC calculations for the $X\,^2\Pi$ state allowed us to verify that the $X\,^2\Pi$ level is the absolute ground state for all molecules considered.

\begin{table*}[tb]
\caption{Computed term values ($\wn$), harmonic vibrational frequencies ($\wn$), anharmonic constants ($\wn$), and  equilibrium bond lengths (\AA) derived from EOMEA-CCSD calculations of Au--group-14 species. Comparisons to experimental data are made, where possible, to the values reported in Ref.~\cite{houdart_emission_1973}.}
\centering
\arrayrulecolor{black}
\begin{tabular*}{0.9\linewidth}{@{\extracolsep{\fill}} cc|cccccccc} 
\toprule
\multicolumn{1}{l}{Molecule} & \multicolumn{1}{l}{State}                               & \multicolumn{2}{c}{$T_e$}                                        & \multicolumn{2}{c}{$\omega_e$}                                 & \multicolumn{2}{c}{$\omega_e x_e$}                           & \multicolumn{2}{c}{$r_e$}   \\ 
\hline
   &                                      & \uline{Comp} & \uline{Exp}                                       & \uline{Comp} & \uline{Exp}                                     & \uline{Comp} & \uline{Exp}                                   & \uline{Comp} & \uline{Exp}  \\
\multirow{3}{*}{AuC}  & $\XPiHalf$                                          & 0            & 0                                                 & 732          & -                                               & 4.4          & -                                             & 1.815        & -            \\
                      & $\XPiThreeHalf$                                     & 2299         & -                                                 & 703          & -                                               & 1.6          & -                                             & 1.841        & -            \\
                     & $\ASigma$                                           & 14475        & -                                                 & 726          & -                                               & 3.5          & -                                             & 1.848        & -            \\
\hline 
 \multirow{3}{*}{AuSi} & $\XPiHalf$                                          & 0            & 0                                                 & 397          & 390.94                                          & 1.5          & 2.22                                          & 2.227        & -            \\
                      & $\XPiThreeHalf$                                     & 1105         & 1072                                              & 385          & -                                               & 1.2          & -                                             & 2.247        & -            \\
                      & $\ASigma$                                           & 12890        & 13632.7                                           & 386          & 389.53                                          & 2.4          & 1.32                                          & 2.248        & -            \\ 
\hline 
\multirow{3}{*}{AuGe} & $\XPiHalf$                                          & 0            & 0                                                 & 265          & 249.68                                          & 0.6          & 0.33                                          & 2.309        & -            \\
                      & $\XPiThreeHalf$                                     & 1460         & 1554                                              & 260          & -                                               & 0.5          & -                                             & 2.324        & -            \\
                      & $\ASigma$                                           & 13013        & 13743.3                                           & 252          & 242.57                                          & 0.8          & 0.59                                          & 2.333        & -            \\ 
\hline
\multirow{3}{*}{AuSn} & $\XPiHalf$                                          & 0            & 0                                                 & 210          & 190.4                                           & 0.9          & 1.26                                          & 2.513        & -            \\
                      & $\XPiThreeHalf$                                     & 2360         & 2551                                              & 209          & -                                               & 0.9          & -                                             & 2.518        & -            \\
                      & $\ASigma$                                           & 13200        & 13899                                             & 193          & 179                                             & 1.7          & 1.44                                          & 2.542        & -            \\ 
\hline
\multirow{3}{*}{AuPb} & $\XPiHalf$                                          & 0            & 0                                                 & 166          & 158.6                                           & 0.3          & 0.6                                           & 2.600        & -            \\
                      & $\XPiThreeHalf$                                     & 7151         & 7500                                              & 178          & -                                               & 0.3          & -                                             & 2.570        & -            \\
                      & {$\ASigma$} & 14680        & 16357.6 & 159          & 152.7 & 0.2          & 0.9 & 2.607        & -            \\
\bottomrule
\end{tabular*}
\arrayrulecolor{black}
\label{tab:AuXpreds}
\end{table*}

To determine vibrational properties and Franck-Condon factors, both the CCSD(T) and EOM-CCSD calculations were performed at a number of Au-X internuclear distances in the vicinity of the equilibrium bond length. These scans produced values along the potential energy curves (PECs) of the low-lying electronic states. The resulting energies were fit to a fourth-order polynomial which were then input to discrete variable representation (DVR) calculations~\cite{tannor_introduction_2008} that compute the vibrational wavefunctions, vibrational energy levels, and Franck-Condon factors. Equilibrium bond lengths were determined from the computed PECs, while harmonic and anharmonic vibrational frequencies were found by fitting the vibrational energy levels to the conventional spectroscopic term values, 
\begin{equation}
    G(v) = \omega_e (v+1/2) - \omega_e x_e (v+1/2)^2.
\end{equation}
The vibrational wavefunctions were used to compute numerical overlap integrals from which Franck-Condon factors were determined. The fitted curves accurately capture the PECs within the limits of our calculation, and the addition of higher-order terms did not significantly alter any numerical results.

\section{Results}

\subsection{Electronic and Fine Structure}

Gold--group 14 dimers have been studied in a limited number of previous experimental and theoretical investigations. AuSi and AuGe were observed as early as 1964 by Barrow and coworkers, who assigned the ground states as $^2\Pi_{1/2}$~\cite{barrow_electronic_1964}. Recent \textit{ab initio} studies have supported this analysis~\cite{barysz_ausi_2021,tran_spinorbit_2018}. Emission spectra of AuSi, AuGe, AuSn, and AuPb were recorded by Houdart and Schamps, allowing vibrational and spin-orbit structure to be observed via optical transitions in the visible and near-infrared regions~\cite{houdart_emission_1973}. 
AuSi has also been the subject of one high-resolution investigation~\cite{scherer_cavity_1995}, though the spectra were assigned to a $\Sigma^+ - \Sigma^+$ band that did not involve the absolute ground state~\cite{boldyrev_ground_1998}. 
Though it has not been studied experimentally, one theoretical investigation of AuC has predicted a $^2\Pi$ ground state~\cite{li_electronic_2011}.

Our computational results are summarized in Tab.~\ref{tab:AuXpreds}. Each AuX molecule possesses a $^2\Pi_{1/2}$ ground state arising from a single unpaired electron in the $2\pi$ orbital (Fig.~\ref{fig:AuC_MOs}). The highest occupied molecular orbital (HOMO) is weakly antibonding and predominantly centered on the carbon-group atom. At the EOMEA-CCSD level of theory, the contribution of Au-centered atomic orbitals to the HOMO decreases as follows: 40\% for AuC, 18\% for AuSi, 14\% for AuGe, and 10\% for AuSn, and 25\% for AuPb. This trend roughly follows the increasing energy of the $n$p atomic orbital on the X atom as one proceeds down the Group 14 elements~\cite{noauthor_atomic_2015}. As the atomic orbital energy increases relative to Au 6s, one would expect to increase the electron density on the X atom. The localization of the radical electron has important implications for the values of $\Eeff$, as discussed below.

Computed values of the molecular dipole moments, $\mu_\mathrm{el}$, and electronic $g$-factors, $g_e$, are reported in Tab.~\ref{tab:Moments}. These parameters are important to electron EDM experiments because $\mu_\mathrm{el}$ partially determines how easily the molecule can be polarized in the laboratory frame and $g_e$ determines the sensitivity of the experiment to magnetic fields. The trend in $\mu_\mathrm{el}$ values reflects the decreasing tendency for the X atom to attract electron density. Interestingly, the observed trend is more uniform than the trend in carbon-group electron affinities. The dipole moment magnitudes indicate that these molecules can be polarized in reasonable laboratory electric fields, as described below. The small values of $g_e$ also support the assignment of the ground state term symbol as $^2\Pi_{1/2}$ due to the near cancellation between the spin and orbital contributions, which is expected for a $^2\Pi_{1/2}$ term. This suggests that mixing with low-lying states of $^2\Sigma$ symmetry is quite suppressed. The very small magnetic sensitivity is beneficial for future EDM experiments using AuX molecules, as will be discussed below.

\begin{table}[tb]
\caption{Electric dipole moments and electronic $g$-factors for the ground state of AuX molecules. Electric dipole moments are reported in units of Debye while $g_e$ is dimensionless.}
\label{tab:Moments}
\setlength\tabcolsep{0pt}
\begin{tabular*}{0.6\linewidth}{@{\extracolsep{\fill}}lcc}
\toprule
Molecule  & $\mu_\mathrm{el}$ & $g_e$  \\ \cline{1-3} 
AuC & $1.92$ & $0.017$  \\
AuSi & $0.21$ & $0.007$  \\
AuGe  & $-0.23$ & $0.003$  \\  
AuSn & $-0.88$ & $0.002$  \\
AuPb & $-1.66$ & $0.055$  \\
\bottomrule 
\end{tabular*}
\end{table}

The $X\,^2\Pi$ state exhibits significant spin-orbit coupling, with \XPiThreeHalf somewhat higher in energy than \XPiHalf. Computed spin-orbit splittings, determined by $T_e(^2\Pi_{3/2})-T_e(^2\Pi_{1/2})$, are in good agreement with previous experimental measurements, as shown in Table~\ref{tab:AuXpreds}~\cite{houdart_emission_1973}. The AuX spin-orbit splitting generally increases with the mass of the group 14 atom, X, indicating a substantial amount of electron density localized on X. The exception of AuC is due to a relatively larger contribution of the Au(6p) orbital to the AuC HOMO. In general, the trend in spin-orbit splittings mirrors the Au-centered electron density that was reported above.

Consistent with the optical spectra reported in Ref.~\cite{houdart_emission_1973}, the lowest-energy doublet excited state in each AuX molecule is $\ASigma$. The $\ASigma$ arises from excitation of the radical electron to the $3\sigma$ MO and displays significant electron density on the Au atom. Despite the significant charge transfer character of this excitation, both the HOMO and LUMO show similar degrees of net antibonding character, which helps to explain the diagonal FCFs that will be discussed in Sec.~\ref{sec:FCFs}. The transition energies predicted by our calculations are in relatively good agreement with measured values~\cite{houdart_emission_1973}, as shown in Tab.~\ref{tab:AuXpreds}. 

Excitation of an electron from the $2\sigma$ orbital to $2\pi$ produces a term of $^4\Sigma^-$ symmetry. A series of SFX2C-1e-ROHF-CCSD(T)/cc-pVTZ calculations predict that, neglecting the spin-orbit interaction, this $a\,^4\Sigma^-$ state is higher than the $X\,^2\Pi$ ground state by approximately 8,500~\wn in AuC, 11,400~\wn in AuSi, 13,600~\wn in AuGe, 13,615~\wn in AuSn, and 14,900~\wn in AuPb. The predictions for AuC and AuSi are similar to previous theoretical studies, which placed the $^4\Sigma^-$ state near, or even above, the $\ASigma$ state~\cite{li_electronic_2011, barysz_ausi_2021, wu_electronic_2006, abe_theoretical_2002}.

\begin{figure*}[tb]
    \centering 
    \includegraphics[width=0.95\textwidth]{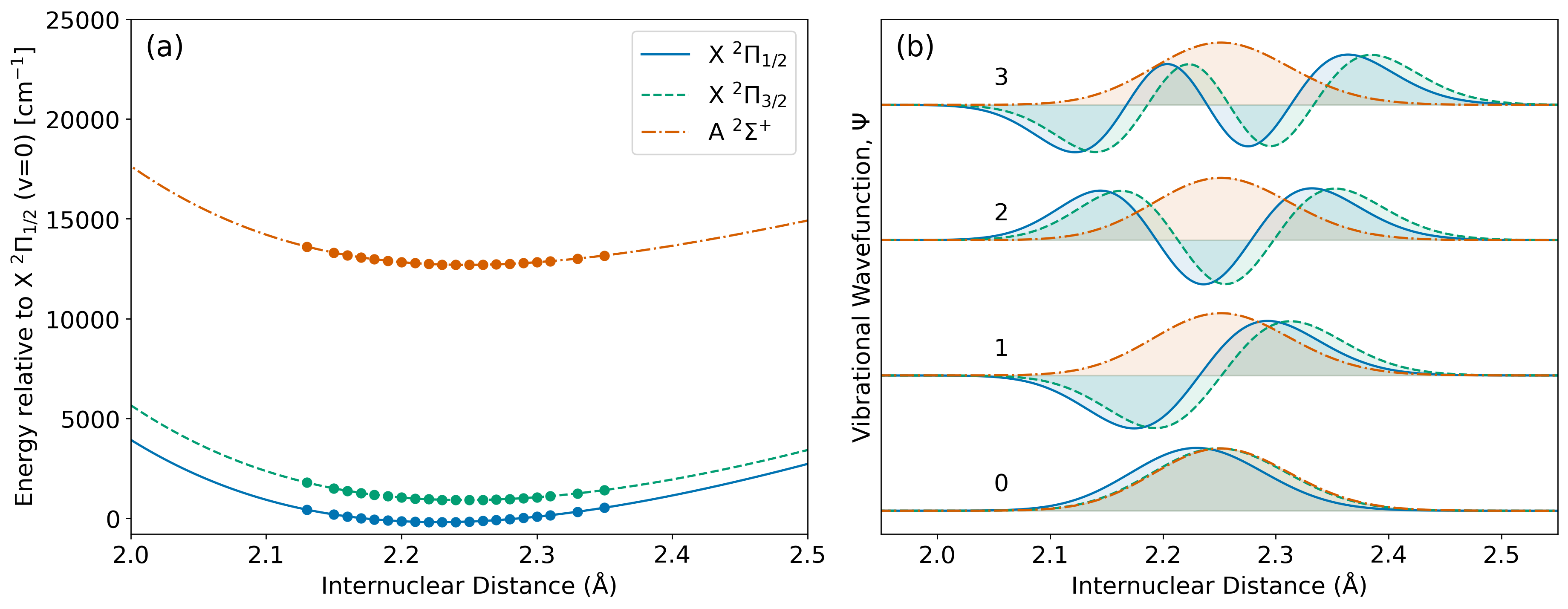}
    \caption{(a) Potential energy curves (PECs) for the lowest doublet states of AuSi, computed using the EOMEA-CCSD single-point energies. Fourth-order polynomial fits to each PEC are also shown. (b) Computed vibrational wavefunctions for $\ASigma(v' = 0)$ (orange, dot-dashed) and the $v''=0, 1, 2, 3$ vibrational levels of \XPiHalf (blue, solid) and \XPiThreeHalf (green, dashed) demonstrate the favorable wavefunction overlap.}
    \label{fig:AuC_PEC}
\end{figure*}

\subsection{Vibrational Structure}

A representative set of PECs computed for AuSi is plotted in Fig.~\ref{fig:AuC_PEC}, and the results of our DVR calculations to obtain vibrational energy levels are reported in Table~\ref{tab:AuXpreds}. The computed vibrational energy levels agree well with experimental measurements~\cite{houdart_emission_1973, scherer_cavity_1995}, suggesting that the potential energy curves and computed vibrational properties model these systems well. Harmonic vibrational constants are generally within about 10~\wn of the measured values, with slightly larger deviations seen in the case of AuSn. Anharmonic constants $\omega_e x_e$ are relatively small ($\lesssim 1$~\wn) and positive. The harmonic vibrational frequencies differ very little between ground and excited states, indicating that the bond strength does not change significantly upon electronic excitation. Inspection of the HOMO and LUMO (Fig.~\ref{fig:AuC_MOs}) helps rationalize this similarity. Both molecular orbitals show a similarly diminished electron density between the nuclei, albeit with significant charge-transfer character in the excitation. Evidently the relative balance of antibonding character does not significantly alter the bond length upon electronic excitation.

\subsection{Vibrational Branching Fractions} \label{sec:FCFs}
The computational results presented above show that the bond lengths and vibrational frequencies of a given AuX species are largely unchanged upon electronic excitation from $\XPiHalf$ or \XPiThreeHalf to $\ASigma$. This suggests that the molecules will exhibit diagonal FCFs, one of the requirements for optical cycling and laser cooling. To support this expectation, we have computed FCFs for the $\ASigma \rightarrow \XPi$ band of each AuX molecule. 

We computed FCFs by constructing overlap integrals between vibrational wavefunctions generated by the DVR calculations. For optical cycling experiments, a more relevant quantity is the vibrational branching fraction (VBF), which represents the relative probability of decay from an excited to a ground vibronic state. To incorporate the presence of both $\Omega = 1/2, 3/2$ components, which in general have different equilibrium bond lengths, we computed separate FCFs for each component. The intrinsic rotational line strengths for decays from $\ASigma(N'=0)$ to $\XPiHalf$ and $\XPiThreeHalf$ are equal. Thus, VBFs are determined by including the effect of different transition frequencies on the spontaneous decay rate according to
\begin{equation}
    b_{v' \rightarrow v} = \frac{\sum_{\Omega}\text{FCF}_{v' \rightarrow (v, \Omega)} \times \nu_{v',(v, \Omega)}^3}{\sum_{v, \Omega} \text{FCF}_{v' \rightarrow (v,\Omega)} \times \nu_{v',(v, \Omega)}^3},
    \label{eq:VBFweighted}
\end{equation}
where $\nu_{v',(v, \Omega)}$ is the transition frequency from $A(v')$ to $X(v, \Omega)$. The sum over $\Omega$ components in this expression produces conventional VBFs, which correspond to a particular $A(v') \rightarrow X(v)$ decay.

\begin{table}[tb]
\caption{Vibrational branching fractions from $\ASigma(v'=0)$ to low-lying vibrational levels of $\XPi$ computed using numerical wavefunctions for AuX molecules. VBFs are derived from EOMEA-CCSD/cc-pVTZ-SO calculations.}
\label{tab:VBRs}
\setlength\tabcolsep{0pt}
\begin{tabular*}{\linewidth}{@{\extracolsep{\fill}}cccccc}
\toprule
  $b_{0\rightarrow v}$      & AuC       & AuSi       & AuGe & AuSn & AuPb \\ \cline{1-6} 
0                           &  0.9347   &  0.9665    &  0.9282    &   0.8429   & 0.9555     \\
1                           &  0.0642   &  0.0323    &  0.0686   &  0.1434    &  0.0388     \\
2                           &  0.0010   &  0.0006    &  0.0031     &  0.0128    &   0.0051    \\
3                           &  $ < 10^{-4}$ & $< 10^{-4}$ & 0.0001    &  0.0009    & 0.0006    \\
\bottomrule
\end{tabular*}
\end{table}

Computed VBFs for the AuX species are presented in Table~\ref{tab:VBRs}. As expected from the small change in equilibrium bond length, our calculated values show that the AuX molecules possess highly diagonal VBFs. The VBFs for the primary optical cycling transition ($\ASigma(v=0) \rightarrow \XPiHalf(v=0)$) range from approximately 0.84 for AuSn to nearly 0.97 for AuSi. These VBF values are similar to, if not higher than, molecules that have already been successfully laser cooled~\cite{barry2014magnetooptical, truppe_molecules_2017,lim2018laser,collopy20183d, augenbraun2020lasercooled, vazquez-carson_direct_2022}.

\subsection{Effective Electric Fields}
The relativistic effects associated with gold's high-$Z$ nucleus~\cite{schwerdtfeger_relativistic_2002,bartlett_relativistic_1998,pyykko_relativistic_2012} make the AuX molecules interesting for precision measurements of fundamental physics, such as searches for the electron electric dipole moment and other time-reversal-violating effects~\cite{byrnes_enhancement_1999}. To assess the utility of AuX molecules for next-generation electron EDM measurements, we have computed the effective electric field, \Eeff, in the absolute ground \XPiHalf state. As described in the Supplemental Information, to validate these calculations, we have used the same computational methods to predict hyperfine coupling constants of related gold-containing molecules and find good agreement with the experimental measurements.

\begin{table}[tb]
\caption{Effective electric field, $|\Eeff|$ (in GV/cm), in the \XPiHalf state of AuX molecules computed at various levels of theory using uncontracted ANO-RCC basis sets.}
\label{tab:Eeff}
\setlength\tabcolsep{0pt}
\begin{tabular*}{\linewidth}{@{\extracolsep{\fill}}lccc}
\toprule
Molecule  & CCSD & CCSD(T) & EOM-CCSD \\ \cline{1-4} 
AuC & 12.2 & 6.8 & 9.3 \\
AuSi & 11.6 & 11.0 & 12.7 \\
AuGe  & 6.9 & 6.3 & 7.8 \\  
AuSn & 1.3 & 1.7 & 1.7 \\
AuPb & 32.3 & 30.8 &  38.5 \\
\bottomrule 
\end{tabular*}
\end{table}

Computed values of $|\Eeff|$ are reported in Table~\ref{tab:Eeff}. The values of $|\Eeff|$ decrease from about 10~GV/cm for AuC and AuSi to 2~GV/cm for AuSn, before increasing to over 30~GV/cm for AuPb. The values are comparable to molecules that have been used in recent limit-setting experiments~\cite{roussy_improved_2023, hudson_improved_2011}.  The trend in $\Eeff$ values can be readily understood based on the compositions of the AuX HOMOs, which were described above. The contribution of Au-centered orbitals to the HOMO decreases from about 40\% to about 10\% as the identity of X progresses down the periodic table from C to Sn. Decreased spin density near the gold nucleus leads to a smaller value of $\Eeff$. The trend in $\Eeff$ values breaks when Au bonds to Pb, because Pb itself has a large nuclear charge, $Z = 82$. AuPb therefore has the largest $\Eeff$ of the AuX series at $>30$~GV/cm. This is similar to the value of $\Eeff$ in other Pb-containing molecules, PbO ($\sim$25~GV/cm in the metastable $a(1)\,^3\Sigma^+$ state~\cite{kozlov_enhancement_2002,eckel_search_2013}) and PbF ($\sim$35~GV/cm in the ground $X\,^2\Pi_{1/2}$ state~\cite{baklanov_progress_2010,skripnikov_search_2014,sasmal_calculation_2015}). This observation suggests that it will be fruitful to investigate whether the other coinage-metal plumbides (CuPb and AgPb) also share the fortuitous combination of diagonal FCFs and large $\Eeff$.

\section{Discussion and Applications}

\begin{figure*}[tb]
    \centering 
    \includegraphics[width=0.9\textwidth]{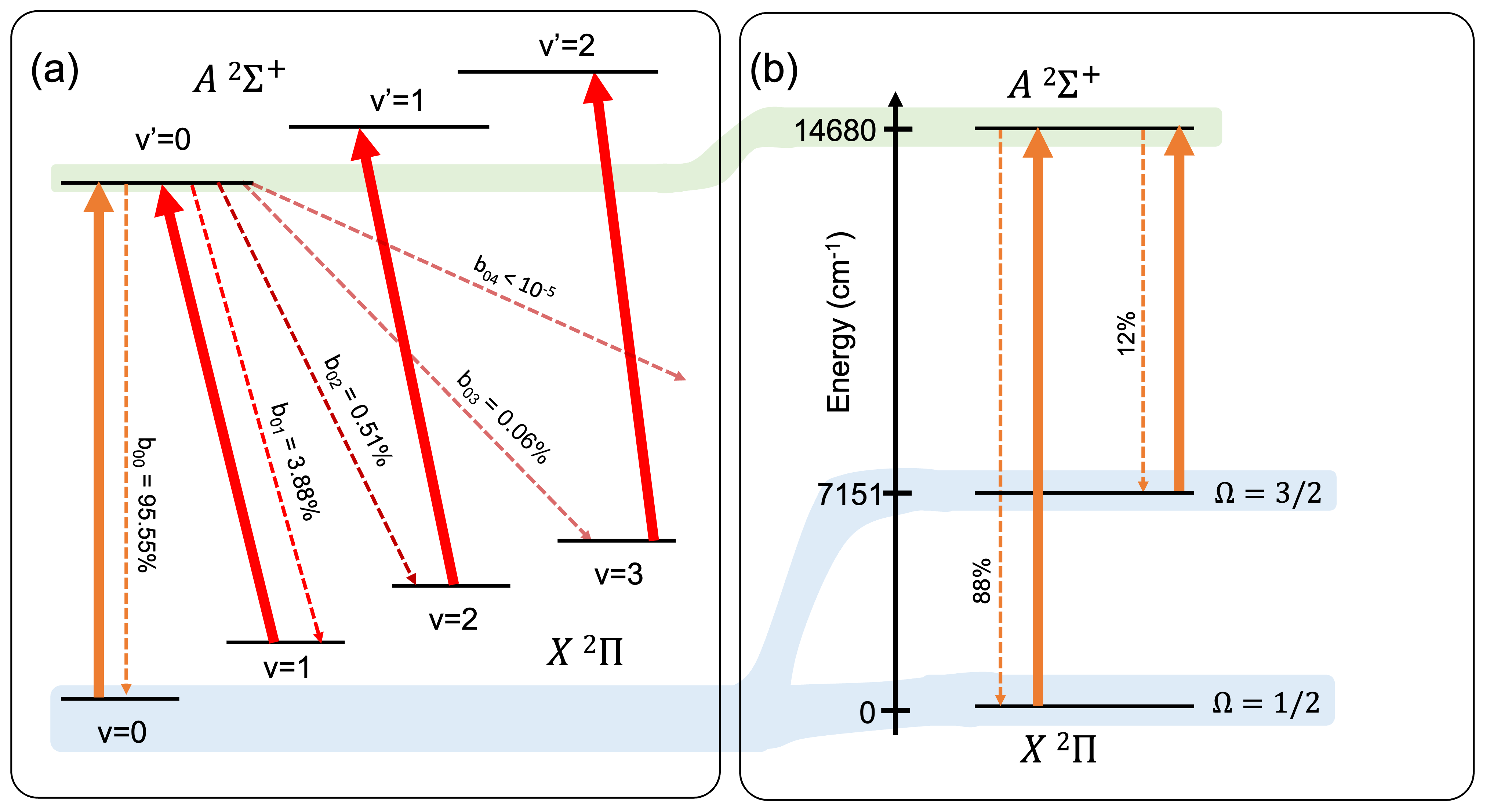}
    \caption{Optical cycling scheme for the example of AuPb. Other AuX molecules will be analogous. (a) Vibrational repumping scheme to achieve near-unity closure of the optical cycle. (b) Fine structure components of each vibrational repumping transition. Calculated rotational branching fractions, computed in a manner analogous to Eq.~\ref{eq:VBFweighted}, are indicated.}
    \label{fig:RotationalScheme}
\end{figure*}

\subsection{Optical Cycling}
Significant work has been devoted to identifying complex molecules that can support optical cycling~\cite{stuhl2008magnetooptical,hamamda2015rovibrational,isaev2016polyatomic,kozyryev2016proposal,augenbraun2020molecular}. The diagonal FCFs predicted for AuX species suggest that these molecules will support optical cycling on the $\ASigma \leftarrow \XPi$ electronic transition. Optical cycling enables unit-efficiency state preparation and non-destructive readout, which is of great utility to electron EDM experiments. It is also a necessary, but not sufficient, ingredient for laser cooling~\cite{fitch2021lasercooleda}. Here, we focus solely on the prospects for optical cycling.

An established method to assess the feasibility of optical cycling is to estimate the minimum number of laser repumpers, $N_L^{5.0}$, needed to keep a molecule within the optical cycle in 99.999\% of all photon scattering events~\cite{fitch2021lasercooleda}. In other words, $N_L^{5.0}$ gives an estimate of the number of lasers required to scatter approximately 100,000 photons per molecule. Table~\ref{tab:FiveNines} presents the degree of optical cycling achievable for each AuX radical. In all cases except AuSn, between 6 and 10 lasers are necessary to achieve a closure of 1 part in $10^5$. Due to spin-orbit coupling, each vibrational repumper requires 2 distinct laser wavelengths, one for the $\XPiHalf$ component and another for the $\XPiThreeHalf$ (see Fig.~\ref{fig:RotationalScheme}). This level of technical complexity is manageable with modern laser systems, and is comparable to that of established optical cycling experiments~\cite{vilas_magneto-optical_2022, zeng_three-dimensional_2024}.  Moreover, the wavelengths required are at technologically convenient wavelengths where solid state lasers can be used. A representative vibrational repumping scheme is shown in Fig.~\ref{fig:RotationalScheme} for the case of AuPb.

Optical cycling on a $^2\Sigma^+ \leftarrow {^2\Pi}$ transition has been previously discussed in the context of laser cooling CH radicals~\cite{schnaubelt2021cold}. Unlike in CH, the sizable spin-orbit coupling in AuX radicals leads to ${}^2\Pi$ states that are well described by a Hund's case (a) basis. A practical optical cycle can be constructed by applying two laser wavelengths per vibronic transition, as shown in Fig.~\ref{fig:RotationalScheme}(b). The predicted hyperfine coupling constants in AuX radicals lead to hyperfine splittings on the order of 100--1000~MHz (see Supplemental Information). These hyperfine splittings will require the addition of frequency sidebands using acousto-optic or electro-optic modulators, as has been demonstrated for molecules with complex hyperfine structure~\cite{zeng_optical_2023}. The precise details of the rotational/hyperfine structure is beyond the scope of this paper and will be the subject of future investigations.

The $a\,^4\Sigma^-$ state is not expected to cause significant loss from the optical cycle. Decays from $\ASigma$ to this intermediate level is forbidden by both spin ($S=1/2 \nrightarrow S=3/2$) and symmetry ($\Sigma^+ \nrightarrow \Sigma^-$) selection rules. Furthermore, the small energy splitting between $\ASigma$ and $a\,^4\Sigma^-$ will strongly suppress the spontaneous emission rate to the quartet state. Future investigations are warranted on the possibility of spin-orbit coupling between the $a\,^4\Sigma^-$ and \ASigma states.

\begin{table}[tb]
\caption{Predicted optical cycling properties of AuX molecules. For each molecule, we list (1) the average number of photons that can be scattered using just 7 repumping lasers and (2) the number of lasers required to achieve an average of 100,000 photons scattered per molecule ($N_L^{5.0}$).}
\label{tab:FiveNines}
\setlength\tabcolsep{0pt}
\begin{tabular*}{\linewidth}{@{\extracolsep{\fill}}lcc}
\toprule
Molecule  & Photons Scattered with 7 repumpers & $N_L^{5.0}$ \\ \cline{1-3} 
AuC & $3.4 \times 10^7$ & 6 \\
AuSi & $4.7 \times 10^6$ & 6 \\
AuGe  & $2.4 \times 10^5$ & 8 \\  
AuSn & $6.6 \times 10^3$ & 14 \\
AuPb & $2.3 \times 10^4$ & 10 \\
\bottomrule 
\end{tabular*}
\end{table}

\subsection{Electron EDM Measurements}
For an experiment using trapped molecules, the shot-noise limited sensitivity to the electron EDM can be estimated as
\begin{equation}
    \delta d_e = \frac{\hbar}{2 \Eeff \sqrt{N \tau} \sqrt{T_\text{int}}},
\end{equation}
where $\Eeff$ is the effective electric field, $N$ is the number of molecules probed per experimental cycle, $\tau$ is the coherence time, and $T_\text{int}$ is the total integration time~\cite{khriplovich_cp_1997}.
As a demonstration of the potential utility of gold-containing molecules for electron EDM measurements, consider an experiment that uses $10^4$ AuPb molecules held in an optical trap with a coherence time of 2~s, all reasonable values given existing experimental platforms. Such an experiment would achieve a statistical sensitivity of $\delta d_e \sim 8 \times 10^{-32}$~$e\cdot$cm with 1 week of averaging. This sensitivity is nearly two orders of magnitude more sensitive than the current limit~\cite{roussy_improved_2023}. 

The $^2\Pi_{1/2}$ ground state conveys several advantages to an electron EDM experiment using AuX molecules. The presence of closely-spaced $\Lambda$-doublets means that the molecules can be easily polarized, providing full access to $\Eeff$ in the laboratory frame. For typical $\Lambda$-doublet splittings of 0.5~GHz (see Supplemental Information) and molecular dipole moments around 2~D, laboratory fields smaller than 1~kV/cm are necessary to fully polarize the molecules. Because the science state is the absolute ground state, it is possible to make use of long spin precession times without becoming limited by spontaneous emission, as would occur in experiments that use metastable states~\cite{kozyryev2017precision}. The $\XPiHalf$ ground state also features $g$-factors that are a factor of at least 50 smaller than that of a ${}^2\Sigma$ electronic state, dramatically reducing the sensitivity of such an experiment to external magnetic fields. At the same time, the $\XPiThreeHalf$ state is also a useful resource to check for systematic errors due to its \textit{large} magnetic sensitivity ($g \approx 2$)~\cite{wu_metastable_2020}. Overall, these molecules appear suitable for sensitive electron EDM measurements, including using recently proposed measurement schemes that rely on ``clock'' states~\cite{verma_electron_2020}.

\section{Conclusion}
We have identified the molecules formed from gold and carbon-group atoms as promising candidates for next-generation electron electric dipole moment measurements. Using high-level electronic structure calculations, we find that these molecules possess nearly diagonal Franck-Condon factors and high intrinsic sensitivity to the electron EDM. The ground state term symbols are identified as $\XPiHalf$, which represent ideal ``science'' states for future precision measurements. Additional theoretical investigations are warranted to study whether AuX species are sensitive to other symmetry-violating electromagnetic moments, such as the nuclear magnetic quadrupole moment. Furthermore, it will be interesting to investigate the electron EDM sensitivity in molecules built from other coinage metals with carbon-group atoms, e.g. CuPb and AgPb.

Our laboratory is pursuing the high-resolution laser spectroscopy to test these theoretical predictions. These measurements will provide insight into the most practical methods to realize cold, trapped samples of AuX molecules, including whether recent ``few-photon" techniques can be readily applied to molecules with ground-state parity doublets~\cite{augenbraun2021zeemansisyphus,fitch2016principles,comparat2014molecular,zeppenfeld2012sisyphus}.  These molecules may represent a new framework with which to construct molecular optical cycling centers, expanding the degree of chemical complexity that can be controlled with laser light for applications in quantum science and precision measurements.

\textit{Supplemental Information ---} Additional computational results and comparisons to experiment.


\section*{Acknowledgments}
We are grateful to Zack D. Lasner (Harvard University) for stimulating discussions, initially identifying AuSi as a molecule of interest, and valuable feedback on this manuscript. We thank Nick Hutzler (Caltech) for useful discussions about this work. BLA acknowledges startup funding from Williams College and support from the ACS Petroleum Research Fund (ACS PRF \#67451-UNI6). Computational work at Williams was conducted using the Williams College High-Performance Computing Cluster. The work at The Johns Hopkins University was supported by the National Science Foundation, under Grant No. PHY-2309253. 


\bibliography{references}

\begin{thebibliography}{93}%
\makeatletter
\providecommand \@ifxundefined [1]{%
 \@ifx{#1\undefined}
}%
\providecommand \@ifnum [1]{%
 \ifnum #1\expandafter \@firstoftwo
 \else \expandafter \@secondoftwo
 \fi
}%
\providecommand \@ifx [1]{%
 \ifx #1\expandafter \@firstoftwo
 \else \expandafter \@secondoftwo
 \fi
}%
\providecommand \natexlab [1]{#1}%
\providecommand \enquote  [1]{``#1''}%
\providecommand \bibnamefont  [1]{#1}%
\providecommand \bibfnamefont [1]{#1}%
\providecommand \citenamefont [1]{#1}%
\providecommand \href@noop [0]{\@secondoftwo}%
\providecommand \href [0]{\begingroup \@sanitize@url \@href}%
\providecommand \@href[1]{\@@startlink{#1}\@@href}%
\providecommand \@@href[1]{\endgroup#1\@@endlink}%
\providecommand \@sanitize@url [0]{\catcode `\\12\catcode `\$12\catcode
  `\&12\catcode `\#12\catcode `\^12\catcode `\_12\catcode `\%12\relax}%
\providecommand \@@startlink[1]{}%
\providecommand \@@endlink[0]{}%
\providecommand \url  [0]{\begingroup\@sanitize@url \@url }%
\providecommand \@url [1]{\endgroup\@href {#1}{\urlprefix }}%
\providecommand \urlprefix  [0]{URL }%
\providecommand \Eprint [0]{\href }%
\providecommand \doibase [0]{https://doi.org/}%
\providecommand \selectlanguage [0]{\@gobble}%
\providecommand \bibinfo  [0]{\@secondoftwo}%
\providecommand \bibfield  [0]{\@secondoftwo}%
\providecommand \translation [1]{[#1]}%
\providecommand \BibitemOpen [0]{}%
\providecommand \bibitemStop [0]{}%
\providecommand \bibitemNoStop [0]{.\EOS\space}%
\providecommand \EOS [0]{\spacefactor3000\relax}%
\providecommand \BibitemShut  [1]{\csname bibitem#1\endcsname}%
\let\auto@bib@innerbib\@empty
\bibitem [{\citenamefont {DeMille}(2015)}]{demille_diatomic_2015}%
  \BibitemOpen
  \bibfield  {author} {\bibinfo {author} {\bibfnamefont {D.}~\bibnamefont
  {DeMille}},\ }\bibfield  {title} {\bibinfo {title} {Diatomic molecules, a
  window onto fundamental physics},\ }\href {https://doi.org/10.1063/PT.3.3020}
  {\bibfield  {journal} {\bibinfo  {journal} {Physics Today}\ }\textbf
  {\bibinfo {volume} {68}},\ \bibinfo {pages} {34} (\bibinfo {year}
  {2015})}\BibitemShut {NoStop}%
\bibitem [{\citenamefont {Safronova}\ \emph {et~al.}(2018)\citenamefont
  {Safronova}, \citenamefont {Budker}, \citenamefont {DeMille}, \citenamefont
  {Kimball}, \citenamefont {Derevianko},\ and\ \citenamefont
  {Clark}}]{safronova2018search}%
  \BibitemOpen
  \bibfield  {author} {\bibinfo {author} {\bibfnamefont {M.~S.}\ \bibnamefont
  {Safronova}}, \bibinfo {author} {\bibfnamefont {D.}~\bibnamefont {Budker}},
  \bibinfo {author} {\bibfnamefont {D.}~\bibnamefont {DeMille}}, \bibinfo
  {author} {\bibfnamefont {D.~F.~J.}\ \bibnamefont {Kimball}}, \bibinfo
  {author} {\bibfnamefont {A.}~\bibnamefont {Derevianko}},\ and\ \bibinfo
  {author} {\bibfnamefont {C.~W.}\ \bibnamefont {Clark}},\ }\bibfield  {title}
  {\bibinfo {title} {Search for new physics with atoms and molecules},\ }\href
  {https://doi.org/10.1103/RevModPhys.90.025008} {\bibfield  {journal}
  {\bibinfo  {journal} {Rev. Mod. Phys.}\ }\textbf {\bibinfo {volume} {90}},\
  \bibinfo {pages} {025008} (\bibinfo {year} {2018})}\BibitemShut {NoStop}%
\bibitem [{\citenamefont {Commins}(2007)}]{commins2007electric}%
  \BibitemOpen
  \bibfield  {author} {\bibinfo {author} {\bibfnamefont {E.~D.}\ \bibnamefont
  {Commins}},\ }\bibfield  {title} {\bibinfo {title} {Electric dipole moments
  of elementary particles, nuclei, atoms, and molecules},\ }\href
  {https://doi.org/10.1143/JPSJ.76.111010} {\bibfield  {journal} {\bibinfo
  {journal} {J. Phys. Soc. Jpn.}\ }\textbf {\bibinfo {volume} {76}},\ \bibinfo
  {pages} {1} (\bibinfo {year} {2007})}\BibitemShut {NoStop}%
\bibitem [{\citenamefont {DeMille}\ \emph {et~al.}(2024)\citenamefont
  {DeMille}, \citenamefont {Hutzler}, \citenamefont {Rey},\ and\ \citenamefont
  {Zelevinsky}}]{demille_quantum_2024}%
  \BibitemOpen
  \bibfield  {author} {\bibinfo {author} {\bibfnamefont {D.}~\bibnamefont
  {DeMille}}, \bibinfo {author} {\bibfnamefont {N.~R.}\ \bibnamefont
  {Hutzler}}, \bibinfo {author} {\bibfnamefont {A.~M.}\ \bibnamefont {Rey}},\
  and\ \bibinfo {author} {\bibfnamefont {T.}~\bibnamefont {Zelevinsky}},\
  }\bibfield  {title} {\bibinfo {title} {Quantum sensing and metrology for
  fundamental physics with molecules},\ }\href
  {https://doi.org/10.1038/s41567-024-02499-9} {\bibfield  {journal} {\bibinfo
  {journal} {Nat. Phys.}\ }\textbf {\bibinfo {volume} {20}},\ \bibinfo {pages}
  {741} (\bibinfo {year} {2024})}\BibitemShut {NoStop}%
\bibitem [{\citenamefont {Roussy}\ \emph {et~al.}(2023)\citenamefont {Roussy},
  \citenamefont {Caldwell}, \citenamefont {Wright}, \citenamefont {Cairncross},
  \citenamefont {Shagam}, \citenamefont {Ng}, \citenamefont {Schlossberger},
  \citenamefont {Park}, \citenamefont {Wang}, \citenamefont {Ye},\ and\
  \citenamefont {Cornell}}]{roussy_improved_2023}%
  \BibitemOpen
  \bibfield  {author} {\bibinfo {author} {\bibfnamefont {T.~S.}\ \bibnamefont
  {Roussy}}, \bibinfo {author} {\bibfnamefont {L.}~\bibnamefont {Caldwell}},
  \bibinfo {author} {\bibfnamefont {T.}~\bibnamefont {Wright}}, \bibinfo
  {author} {\bibfnamefont {W.~B.}\ \bibnamefont {Cairncross}}, \bibinfo
  {author} {\bibfnamefont {Y.}~\bibnamefont {Shagam}}, \bibinfo {author}
  {\bibfnamefont {K.~B.}\ \bibnamefont {Ng}}, \bibinfo {author} {\bibfnamefont
  {N.}~\bibnamefont {Schlossberger}}, \bibinfo {author} {\bibfnamefont {S.~Y.}\
  \bibnamefont {Park}}, \bibinfo {author} {\bibfnamefont {A.}~\bibnamefont
  {Wang}}, \bibinfo {author} {\bibfnamefont {J.}~\bibnamefont {Ye}},\ and\
  \bibinfo {author} {\bibfnamefont {E.~A.}\ \bibnamefont {Cornell}},\
  }\bibfield  {title} {\bibinfo {title} {An improved bound on the electron’s
  electric dipole moment},\ }\href {https://doi.org/10.1126/science.adg4084}
  {\bibfield  {journal} {\bibinfo  {journal} {Science}\ }\textbf {\bibinfo
  {volume} {381}},\ \bibinfo {pages} {46} (\bibinfo {year} {2023})}\BibitemShut
  {NoStop}%
\bibitem [{\citenamefont {{ACME
  Collaboration}}(2018)}]{acme_collaboration_improved_2018}%
  \BibitemOpen
  \bibfield  {author} {\bibinfo {author} {\bibnamefont {{ACME
  Collaboration}}},\ }\bibfield  {title} {\bibinfo {title} {Improved limit on
  the electric dipole moment of the electron},\ }\href
  {https://doi.org/10.1038/s41586-018-0599-8} {\bibfield  {journal} {\bibinfo
  {journal} {Nature}\ }\textbf {\bibinfo {volume} {562}},\ \bibinfo {pages}
  {355} (\bibinfo {year} {2018})}\BibitemShut {NoStop}%
\bibitem [{\citenamefont {Meyer}\ \emph {et~al.}(2006)\citenamefont {Meyer},
  \citenamefont {Bohn},\ and\ \citenamefont
  {Deskevich}}]{meyer_candidate_2006}%
  \BibitemOpen
  \bibfield  {author} {\bibinfo {author} {\bibfnamefont {E.~R.}\ \bibnamefont
  {Meyer}}, \bibinfo {author} {\bibfnamefont {J.~L.}\ \bibnamefont {Bohn}},\
  and\ \bibinfo {author} {\bibfnamefont {M.~P.}\ \bibnamefont {Deskevich}},\
  }\bibfield  {title} {\bibinfo {title} {Candidate molecular ions for an
  electron electric dipole moment experiment},\ }\href
  {https://doi.org/10.1103/PhysRevA.73.062108} {\bibfield  {journal} {\bibinfo
  {journal} {Phys. Rev. A}\ }\textbf {\bibinfo {volume} {73}},\ \bibinfo
  {pages} {062108} (\bibinfo {year} {2006})}\BibitemShut {NoStop}%
\bibitem [{\citenamefont {Eckel}\ \emph {et~al.}(2013)\citenamefont {Eckel},
  \citenamefont {Hamilton}, \citenamefont {Kirilov}, \citenamefont {Smith},\
  and\ \citenamefont {DeMille}}]{eckel_search_2013}%
  \BibitemOpen
  \bibfield  {author} {\bibinfo {author} {\bibfnamefont {S.}~\bibnamefont
  {Eckel}}, \bibinfo {author} {\bibfnamefont {P.}~\bibnamefont {Hamilton}},
  \bibinfo {author} {\bibfnamefont {E.}~\bibnamefont {Kirilov}}, \bibinfo
  {author} {\bibfnamefont {H.~W.}\ \bibnamefont {Smith}},\ and\ \bibinfo
  {author} {\bibfnamefont {D.}~\bibnamefont {DeMille}},\ }\bibfield  {title}
  {\bibinfo {title} {Search for the electron electric dipole moment using
  {$\Omega$}-doublet levels in {PbO}},\ }\href
  {https://doi.org/10.1103/PhysRevA.87.052130} {\bibfield  {journal} {\bibinfo
  {journal} {Phys. Rev. A}\ }\textbf {\bibinfo {volume} {87}},\ \bibinfo
  {pages} {052130} (\bibinfo {year} {2013})}\BibitemShut {NoStop}%
\bibitem [{\citenamefont {Hutzler}(2020)}]{hutzler_polyatomic_2020}%
  \BibitemOpen
  \bibfield  {author} {\bibinfo {author} {\bibfnamefont {N.~R.}\ \bibnamefont
  {Hutzler}},\ }\bibfield  {title} {\bibinfo {title} {Polyatomic molecules as
  quantum sensors for fundamental physics},\ }\href
  {https://doi.org/10.1088/2058-9565/abb9c5} {\bibfield  {journal} {\bibinfo
  {journal} {Quantum Sci. Technol.}\ }\textbf {\bibinfo {volume} {5}},\
  \bibinfo {pages} {044011} (\bibinfo {year} {2020})}\BibitemShut {NoStop}%
\bibitem [{\citenamefont {{The ACME Collaboration}}\ \emph
  {et~al.}(2014)\citenamefont {{The ACME Collaboration}}, \citenamefont
  {Baron}, \citenamefont {Campbell}, \citenamefont {DeMille}, \citenamefont
  {Doyle}, \citenamefont {Gabrielse}, \citenamefont {Gurevich}, \citenamefont
  {Hess}, \citenamefont {Hutzler}, \citenamefont {Kirilov}, \citenamefont
  {Kozyryev}, \citenamefont {O’Leary}, \citenamefont {Panda}, \citenamefont
  {Parsons}, \citenamefont {Petrik}, \citenamefont {Spaun}, \citenamefont
  {Vutha},\ and\ \citenamefont {West}}]{the_acme_collaboration_order_2014}%
  \BibitemOpen
  \bibfield  {author} {\bibinfo {author} {\bibnamefont {{The ACME
  Collaboration}}}, \bibinfo {author} {\bibfnamefont {J.}~\bibnamefont
  {Baron}}, \bibinfo {author} {\bibfnamefont {W.~C.}\ \bibnamefont {Campbell}},
  \bibinfo {author} {\bibfnamefont {D.}~\bibnamefont {DeMille}}, \bibinfo
  {author} {\bibfnamefont {J.~M.}\ \bibnamefont {Doyle}}, \bibinfo {author}
  {\bibfnamefont {G.}~\bibnamefont {Gabrielse}}, \bibinfo {author}
  {\bibfnamefont {Y.~V.}\ \bibnamefont {Gurevich}}, \bibinfo {author}
  {\bibfnamefont {P.~W.}\ \bibnamefont {Hess}}, \bibinfo {author}
  {\bibfnamefont {N.~R.}\ \bibnamefont {Hutzler}}, \bibinfo {author}
  {\bibfnamefont {E.}~\bibnamefont {Kirilov}}, \bibinfo {author} {\bibfnamefont
  {I.}~\bibnamefont {Kozyryev}}, \bibinfo {author} {\bibfnamefont {B.~R.}\
  \bibnamefont {O’Leary}}, \bibinfo {author} {\bibfnamefont {C.~D.}\
  \bibnamefont {Panda}}, \bibinfo {author} {\bibfnamefont {M.~F.}\ \bibnamefont
  {Parsons}}, \bibinfo {author} {\bibfnamefont {E.~S.}\ \bibnamefont {Petrik}},
  \bibinfo {author} {\bibfnamefont {B.}~\bibnamefont {Spaun}}, \bibinfo
  {author} {\bibfnamefont {A.~C.}\ \bibnamefont {Vutha}},\ and\ \bibinfo
  {author} {\bibfnamefont {A.~D.}\ \bibnamefont {West}},\ }\bibfield  {title}
  {\bibinfo {title} {Order of {Magnitude} {Smaller} {Limit} on the {Electric}
  {Dipole} {Moment} of the {Electron}},\ }\href
  {https://doi.org/10.1126/science.1248213} {\bibfield  {journal} {\bibinfo
  {journal} {Science}\ }\textbf {\bibinfo {volume} {343}},\ \bibinfo {pages}
  {269} (\bibinfo {year} {2014})}\BibitemShut {NoStop}%
\bibitem [{\citenamefont {Cairncross}\ \emph {et~al.}(2017)\citenamefont
  {Cairncross}, \citenamefont {Gresh}, \citenamefont {Grau}, \citenamefont
  {Cossel}, \citenamefont {Roussy}, \citenamefont {Ni}, \citenamefont {Zhou},
  \citenamefont {Ye},\ and\ \citenamefont
  {Cornell}}]{cairncross_precision_2017}%
  \BibitemOpen
  \bibfield  {author} {\bibinfo {author} {\bibfnamefont {W.~B.}\ \bibnamefont
  {Cairncross}}, \bibinfo {author} {\bibfnamefont {D.~N.}\ \bibnamefont
  {Gresh}}, \bibinfo {author} {\bibfnamefont {M.}~\bibnamefont {Grau}},
  \bibinfo {author} {\bibfnamefont {K.~C.}\ \bibnamefont {Cossel}}, \bibinfo
  {author} {\bibfnamefont {T.~S.}\ \bibnamefont {Roussy}}, \bibinfo {author}
  {\bibfnamefont {Y.}~\bibnamefont {Ni}}, \bibinfo {author} {\bibfnamefont
  {Y.}~\bibnamefont {Zhou}}, \bibinfo {author} {\bibfnamefont {J.}~\bibnamefont
  {Ye}},\ and\ \bibinfo {author} {\bibfnamefont {E.~A.}\ \bibnamefont
  {Cornell}},\ }\bibfield  {title} {\bibinfo {title} {Precision {Measurement}
  of the {Electron}'s {Electric} {Dipole} {Moment} {Using} {Trapped}
  {Molecular} {Ions}},\ }\href {https://doi.org/10.1103/PhysRevLett.119.153001}
  {\bibfield  {journal} {\bibinfo  {journal} {Phys. Rev. Lett.}\ }\textbf
  {\bibinfo {volume} {119}},\ \bibinfo {pages} {153001} (\bibinfo {year}
  {2017})}\BibitemShut {NoStop}%
\bibitem [{\citenamefont {McCarron}(2018)}]{mccarron_laser_2018}%
  \BibitemOpen
  \bibfield  {author} {\bibinfo {author} {\bibfnamefont {D.}~\bibnamefont
  {McCarron}},\ }\bibfield  {title} {\bibinfo {title} {Laser cooling and
  trapping molecules},\ }\href {https://doi.org/10.1088/1361-6455/aadfba}
  {\bibfield  {journal} {\bibinfo  {journal} {J. Phys. B: At. Mol. Opt. Phys.}\
  }\textbf {\bibinfo {volume} {51}},\ \bibinfo {pages} {212001} (\bibinfo
  {year} {2018})}\BibitemShut {NoStop}%
\bibitem [{\citenamefont {Fitch}\ and\ \citenamefont
  {Tarbutt}(2021)}]{fitch2021lasercooleda}%
  \BibitemOpen
  \bibfield  {author} {\bibinfo {author} {\bibfnamefont {N.}~\bibnamefont
  {Fitch}}\ and\ \bibinfo {author} {\bibfnamefont {M.}~\bibnamefont
  {Tarbutt}},\ }\bibfield  {title} {\bibinfo {title} {Laser-cooled molecules},\
  }in\ \href {https://doi.org/10.1016/bs.aamop.2021.04.003} {\emph {\bibinfo
  {booktitle} {Advances {In} {Atomic}, {Molecular}, and {Optical}
  {Physics}}}},\ Vol.~\bibinfo {volume} {70}\ (\bibinfo  {publisher}
  {Elsevier},\ \bibinfo {year} {2021})\ pp.\ \bibinfo {pages}
  {157--262}\BibitemShut {NoStop}%
\bibitem [{\citenamefont {Fitch}\ \emph {et~al.}(2020)\citenamefont {Fitch},
  \citenamefont {Lim}, \citenamefont {Hinds}, \citenamefont {Sauer},\ and\
  \citenamefont {Tarbutt}}]{fitch_methods_2020}%
  \BibitemOpen
  \bibfield  {author} {\bibinfo {author} {\bibfnamefont {N.~J.}\ \bibnamefont
  {Fitch}}, \bibinfo {author} {\bibfnamefont {J.}~\bibnamefont {Lim}}, \bibinfo
  {author} {\bibfnamefont {E.~A.}\ \bibnamefont {Hinds}}, \bibinfo {author}
  {\bibfnamefont {B.~E.}\ \bibnamefont {Sauer}},\ and\ \bibinfo {author}
  {\bibfnamefont {M.~R.}\ \bibnamefont {Tarbutt}},\ }\bibfield  {title}
  {\bibinfo {title} {Methods for measuring the electron’s electric dipole
  moment using ultracold {YbF} molecules},\ }\href
  {https://doi.org/10.1088/2058-9565/abc931} {\bibfield  {journal} {\bibinfo
  {journal} {Quantum Sci. Technol.}\ }\textbf {\bibinfo {volume} {6}},\
  \bibinfo {pages} {014006} (\bibinfo {year} {2020})}\BibitemShut {NoStop}%
\bibitem [{\citenamefont {Smallman}\ \emph {et~al.}(2014)\citenamefont
  {Smallman}, \citenamefont {Wang}, \citenamefont {Steimle}, \citenamefont
  {Tarbutt},\ and\ \citenamefont {Hinds}}]{smallman_radiative_2014}%
  \BibitemOpen
  \bibfield  {author} {\bibinfo {author} {\bibfnamefont {I.~J.}\ \bibnamefont
  {Smallman}}, \bibinfo {author} {\bibfnamefont {F.}~\bibnamefont {Wang}},
  \bibinfo {author} {\bibfnamefont {T.~C.}\ \bibnamefont {Steimle}}, \bibinfo
  {author} {\bibfnamefont {M.~R.}\ \bibnamefont {Tarbutt}},\ and\ \bibinfo
  {author} {\bibfnamefont {E.~A.}\ \bibnamefont {Hinds}},\ }\bibfield  {title}
  {\bibinfo {title} {Radiative branching ratios for excited states of
  $^{174}${YbF}: {A}pplication to laser cooling},\ }\href
  {https://doi.org/10.1016/j.jms.2014.02.006} {\bibfield  {journal} {\bibinfo
  {journal} {J. Mol. Spec.}\ }\textbf {\bibinfo {volume} {300}},\ \bibinfo
  {pages} {3} (\bibinfo {year} {2014})}\BibitemShut {NoStop}%
\bibitem [{\citenamefont {Chen}\ \emph {et~al.}(2016)\citenamefont {Chen},
  \citenamefont {Bu},\ and\ \citenamefont {Yan}}]{chen_structure_2016}%
  \BibitemOpen
  \bibfield  {author} {\bibinfo {author} {\bibfnamefont {T.}~\bibnamefont
  {Chen}}, \bibinfo {author} {\bibfnamefont {W.}~\bibnamefont {Bu}},\ and\
  \bibinfo {author} {\bibfnamefont {B.}~\bibnamefont {Yan}},\ }\bibfield
  {title} {\bibinfo {title} {Structure, branching ratios, and a laser-cooling
  scheme for the $^{138}${BaF} molecule},\ }\href
  {https://doi.org/10.1103/PhysRevA.94.063415} {\bibfield  {journal} {\bibinfo
  {journal} {Phys. Rev. A}\ }\textbf {\bibinfo {volume} {94}},\ \bibinfo
  {pages} {063415} (\bibinfo {year} {2016})}\BibitemShut {NoStop}%
\bibitem [{\citenamefont {Isaev}\ \emph {et~al.}(2010)\citenamefont {Isaev},
  \citenamefont {Hoekstra},\ and\ \citenamefont
  {Berger}}]{isaev_laser-cooled_2010}%
  \BibitemOpen
  \bibfield  {author} {\bibinfo {author} {\bibfnamefont {T.~A.}\ \bibnamefont
  {Isaev}}, \bibinfo {author} {\bibfnamefont {S.}~\bibnamefont {Hoekstra}},\
  and\ \bibinfo {author} {\bibfnamefont {R.}~\bibnamefont {Berger}},\
  }\bibfield  {title} {\bibinfo {title} {Laser-cooled {RaF} as a promising
  candidate to measure molecular parity violation},\ }\href
  {https://doi.org/10.1103/PhysRevA.82.052521} {\bibfield  {journal} {\bibinfo
  {journal} {Phys. Rev. A}\ }\textbf {\bibinfo {volume} {82}},\ \bibinfo
  {pages} {052521} (\bibinfo {year} {2010})}\BibitemShut {NoStop}%
\bibitem [{\citenamefont {Ng}\ \emph {et~al.}(2022)\citenamefont {Ng},
  \citenamefont {Zhou}, \citenamefont {Cheng}, \citenamefont {Schlossberger},
  \citenamefont {Park}, \citenamefont {Roussy}, \citenamefont {Caldwell},
  \citenamefont {Shagam}, \citenamefont {Vigil}, \citenamefont {Cornell},\ and\
  \citenamefont {Ye}}]{ng_spectroscopy_2022}%
  \BibitemOpen
  \bibfield  {author} {\bibinfo {author} {\bibfnamefont {K.~B.}\ \bibnamefont
  {Ng}}, \bibinfo {author} {\bibfnamefont {Y.}~\bibnamefont {Zhou}}, \bibinfo
  {author} {\bibfnamefont {L.}~\bibnamefont {Cheng}}, \bibinfo {author}
  {\bibfnamefont {N.}~\bibnamefont {Schlossberger}}, \bibinfo {author}
  {\bibfnamefont {S.~Y.}\ \bibnamefont {Park}}, \bibinfo {author}
  {\bibfnamefont {T.~S.}\ \bibnamefont {Roussy}}, \bibinfo {author}
  {\bibfnamefont {L.}~\bibnamefont {Caldwell}}, \bibinfo {author}
  {\bibfnamefont {Y.}~\bibnamefont {Shagam}}, \bibinfo {author} {\bibfnamefont
  {A.~J.}\ \bibnamefont {Vigil}}, \bibinfo {author} {\bibfnamefont {E.~A.}\
  \bibnamefont {Cornell}},\ and\ \bibinfo {author} {\bibfnamefont
  {J.}~\bibnamefont {Ye}},\ }\bibfield  {title} {\bibinfo {title} {Spectroscopy
  on the electron-electric-dipole-moment--sensitive states of {ThF}$^+$},\
  }\href {https://doi.org/10.1103/PhysRevA.105.022823} {\bibfield  {journal}
  {\bibinfo  {journal} {Phys. Rev. A}\ }\textbf {\bibinfo {volume} {105}},\
  \bibinfo {pages} {022823} (\bibinfo {year} {2022})}\BibitemShut {NoStop}%
\bibitem [{\citenamefont {Kozyryev}\ and\ \citenamefont
  {Hutzler}(2017)}]{kozyryev2017precision}%
  \BibitemOpen
  \bibfield  {author} {\bibinfo {author} {\bibfnamefont {I.}~\bibnamefont
  {Kozyryev}}\ and\ \bibinfo {author} {\bibfnamefont {N.~R.}\ \bibnamefont
  {Hutzler}},\ }\bibfield  {title} {\bibinfo {title} {Precision measurement of
  time-reversal symmetry violation with laser-cooled polyatomic molecules},\
  }\href {https://doi.org/10.1103/PhysRevLett.119.133002} {\bibfield  {journal}
  {\bibinfo  {journal} {Phys. Rev. Lett.}\ }\textbf {\bibinfo {volume} {119}},\
  \bibinfo {pages} {133002} (\bibinfo {year} {2017})}\BibitemShut {NoStop}%
\bibitem [{\citenamefont {Stanton}\ \emph {et~al.}()\citenamefont {Stanton},
  \citenamefont {Gauss}, \citenamefont {Cheng}, \citenamefont {Harding},
  \citenamefont {Matthews},\ and\ \citenamefont {Szalay}}]{CFOUR}%
  \BibitemOpen
  \bibfield  {author} {\bibinfo {author} {\bibfnamefont {J.~F.}\ \bibnamefont
  {Stanton}}, \bibinfo {author} {\bibfnamefont {J.}~\bibnamefont {Gauss}},
  \bibinfo {author} {\bibfnamefont {L.}~\bibnamefont {Cheng}}, \bibinfo
  {author} {\bibfnamefont {M.~E.}\ \bibnamefont {Harding}}, \bibinfo {author}
  {\bibfnamefont {D.~A.}\ \bibnamefont {Matthews}},\ and\ \bibinfo {author}
  {\bibfnamefont {P.~G.}\ \bibnamefont {Szalay}},\ }\href@noop {} {\bibinfo
  {title} {{CFOUR, Coupled-Cluster techniques for Computational Chemistry, a
  quantum-chemical program package}}},\ \bibinfo {note} {{W}ith contributions
  from {A}.{A}. {A}uer, {A}. {A}sthana, {R}.{J}. {B}artlett, {U}. {B}enedikt,
  {C}. {B}erger, {D}.{E}. {B}ernholdt, {S.} {B}laschke, {Y}. {J}. {B}omble,
  {S.} {B}urger, {O}. {C}hristiansen, {D.} Datta, {F}. Engel, {R}. Faber, {J.}
  {G}reiner, {M}. {H}eckert, {O}. {H}eun, {M}. Hilgenberg, {C}. {H}uber,
  {T}.-{C}. {J}agau, {D}. {J}onsson, {J}. {J}us{\'e}lius, {T}. Kirsch, {K}.
  {K}lein, {G}.{M.} Kopper, {W}.{J}. {L}auderdale, {F}. {L}ipparini, {J}.
  {L}iu, {T}. {M}etzroth, {L}.{A}. {M}{\"u}ck, {D}.{P}. {O}'{N}eill, {T.}
  {N}ottoli, {D}.{R}. {P}rice, {E}. {P}rochnow, {C}. {P}uzzarini, {K}. {R}uud,
  {F}. {S}chiffmann, {W}. {S}chwalbach, {C}. {S}immons, {S}. {S}topkowicz, {A}.
  {T}ajti, {J}. {V}{\'a}zquez, {F}. {W}ang, {X}. {W}ang, {J}.{D}. {W}atts, {C}.
  {Z}hang, {X}. {Z}heng, and the integral packages {MOLECULE} ({J}.
  {A}lml{\"o}f and {P}.{R}. {T}aylor), {PROPS} ({P}.{R}. {T}aylor), {ABACUS}
  ({T}. {H}elgaker, {H}.{J}. {A}a. {J}ensen, {P}. {J}{\o}rgensen, and {J}.
  {O}lsen), and {ECP} routines by {A}. {V}. {M}itin and {C}. van {W}{\"u}llen.
  {F}or the current version, see http://www.cfour.de.}\BibitemShut {Stop}%
\bibitem [{\citenamefont {Matthews}\ \emph {et~al.}(2020)\citenamefont
  {Matthews}, \citenamefont {Cheng}, \citenamefont {Harding}, \citenamefont
  {Lipparini}, \citenamefont {Stopkowicz}, \citenamefont {Jagau}, \citenamefont
  {Szalay}, \citenamefont {Gauss},\ and\ \citenamefont
  {Stanton}}]{Matthews2020}%
  \BibitemOpen
  \bibfield  {author} {\bibinfo {author} {\bibfnamefont {D.~A.}\ \bibnamefont
  {Matthews}}, \bibinfo {author} {\bibfnamefont {L.}~\bibnamefont {Cheng}},
  \bibinfo {author} {\bibfnamefont {M.~E.}\ \bibnamefont {Harding}}, \bibinfo
  {author} {\bibfnamefont {F.}~\bibnamefont {Lipparini}}, \bibinfo {author}
  {\bibfnamefont {S.}~\bibnamefont {Stopkowicz}}, \bibinfo {author}
  {\bibfnamefont {T.~C.}\ \bibnamefont {Jagau}}, \bibinfo {author}
  {\bibfnamefont {P.~G.}\ \bibnamefont {Szalay}}, \bibinfo {author}
  {\bibfnamefont {J.}~\bibnamefont {Gauss}},\ and\ \bibinfo {author}
  {\bibfnamefont {J.~F.}\ \bibnamefont {Stanton}},\ }\bibfield  {title}
  {\bibinfo {title} {Coupled-cluster techniques for computational chemistry:
  {The} {CFOUR} program package},\ }\href {https://doi.org/10.1063/5.0004837}
  {\bibfield  {journal} {\bibinfo  {journal} {J. Chem. Phys.}\ }\textbf
  {\bibinfo {volume} {152}},\ \bibinfo {pages} {214108} (\bibinfo {year}
  {2020})}\BibitemShut {NoStop}%
\bibitem [{\citenamefont {Stanton}\ \emph {et~al.}(1991)\citenamefont
  {Stanton}, \citenamefont {Gauss}, \citenamefont {Watts},\ and\ \citenamefont
  {Bartlett}}]{Stanton91a}%
  \BibitemOpen
  \bibfield  {author} {\bibinfo {author} {\bibfnamefont {J.~F.}\ \bibnamefont
  {Stanton}}, \bibinfo {author} {\bibfnamefont {J.}~\bibnamefont {Gauss}},
  \bibinfo {author} {\bibfnamefont {J.~D.}\ \bibnamefont {Watts}},\ and\
  \bibinfo {author} {\bibfnamefont {R.~J.}\ \bibnamefont {Bartlett}},\
  }\bibfield  {title} {\bibinfo {title} {A direct product decomposition
  approach for symmetry exploitation in many-body methods. {I. Energy}
  calculation},\ }\href {https://doi.org/10.1063/1.460620} {\bibfield
  {journal} {\bibinfo  {journal} {J. Chem. Phys.}\ }\textbf {\bibinfo {volume}
  {94}},\ \bibinfo {pages} {4334} (\bibinfo {year} {1991})}\BibitemShut
  {NoStop}%
\bibitem [{\citenamefont {Liu}\ \emph {et~al.}(2018)\citenamefont {Liu},
  \citenamefont {Shen}, \citenamefont {Asthana},\ and\ \citenamefont
  {Cheng}}]{Liu18b}%
  \BibitemOpen
  \bibfield  {author} {\bibinfo {author} {\bibfnamefont {J.}~\bibnamefont
  {Liu}}, \bibinfo {author} {\bibfnamefont {Y.}~\bibnamefont {Shen}}, \bibinfo
  {author} {\bibfnamefont {A.}~\bibnamefont {Asthana}},\ and\ \bibinfo {author}
  {\bibfnamefont {L.}~\bibnamefont {Cheng}},\ }\bibfield  {title} {\bibinfo
  {title} {Two-component relativistic coupled-cluster methods using mean-field
  spin-orbit integrals},\ }\href {https://doi.org/10.1063/1.5009177} {\bibfield
   {journal} {\bibinfo  {journal} {J. Chem. Phys.}\ }\textbf {\bibinfo {volume}
  {148}},\ \bibinfo {pages} {034106} (\bibinfo {year} {2018})}\BibitemShut
  {NoStop}%
\bibitem [{\citenamefont {Dyall}(1997)}]{Dyall97}%
  \BibitemOpen
  \bibfield  {author} {\bibinfo {author} {\bibfnamefont {K.~G.}\ \bibnamefont
  {Dyall}},\ }\bibfield  {title} {\bibinfo {title} {{Interfacing relativistic
  and nonrelativistic methods. I. Normalized elimination of the small component
  in the modified Dirac equation}},\ }\href {https://doi.org/10.1063/1.473860}
  {\bibfield  {journal} {\bibinfo  {journal} {J. Chem. Phys.}\ }\textbf
  {\bibinfo {volume} {106}},\ \bibinfo {pages} {9618} (\bibinfo {year}
  {1997})}\BibitemShut {NoStop}%
\bibitem [{\citenamefont {Kutzelnigg}\ and\ \citenamefont
  {Liu}(2005)}]{Kutzelnigg05}%
  \BibitemOpen
  \bibfield  {author} {\bibinfo {author} {\bibfnamefont {W.}~\bibnamefont
  {Kutzelnigg}}\ and\ \bibinfo {author} {\bibfnamefont {W.}~\bibnamefont
  {Liu}},\ }\bibfield  {title} {\bibinfo {title} {{Quasirelativistic theory
  equivalent to fully relativistic theory}},\ }\href
  {https://doi.org/10.1063/1.2137315} {\bibfield  {journal} {\bibinfo
  {journal} {J. Chem. Phys.}\ }\textbf {\bibinfo {volume} {123}},\ \bibinfo
  {pages} {241102} (\bibinfo {year} {2005})}\BibitemShut {NoStop}%
\bibitem [{\citenamefont {Liu}\ and\ \citenamefont {Peng}(2009)}]{Liu09}%
  \BibitemOpen
  \bibfield  {author} {\bibinfo {author} {\bibfnamefont {W.}~\bibnamefont
  {Liu}}\ and\ \bibinfo {author} {\bibfnamefont {D.}~\bibnamefont {Peng}},\
  }\bibfield  {title} {\bibinfo {title} {Exact two-component {Hamiltonians}
  revisited},\ }\href {https://doi.org/10.1063/1.3159445} {\bibfield  {journal}
  {\bibinfo  {journal} {J. Chem. Phys.}\ }\textbf {\bibinfo {volume} {131}},\
  \bibinfo {pages} {031104} (\bibinfo {year} {2009})}\BibitemShut {NoStop}%
\bibitem [{\citenamefont {He{\ss}}\ \emph {et~al.}(1996)\citenamefont
  {He{\ss}}, \citenamefont {Marian}, \citenamefont {Wahlgren},\ and\
  \citenamefont {Gropen}}]{Hess96a}%
  \BibitemOpen
  \bibfield  {author} {\bibinfo {author} {\bibfnamefont {B.~A.}\ \bibnamefont
  {He{\ss}}}, \bibinfo {author} {\bibfnamefont {C.~M.}\ \bibnamefont {Marian}},
  \bibinfo {author} {\bibfnamefont {U.}~\bibnamefont {Wahlgren}},\ and\
  \bibinfo {author} {\bibfnamefont {O.}~\bibnamefont {Gropen}},\ }\bibfield
  {title} {\bibinfo {title} {{A mean-field spin-orbit method applicable to
  correlated wavefunctions}},\ }\href
  {https://doi.org/10.1016/0009-2614(96)00119-4} {\bibfield  {journal}
  {\bibinfo  {journal} {Chem. Phys. Lett.}\ }\textbf {\bibinfo {volume}
  {251}},\ \bibinfo {pages} {365} (\bibinfo {year} {1996})}\BibitemShut
  {NoStop}%
\bibitem [{\citenamefont {Liu}\ and\ \citenamefont
  {Cheng}(2018)}]{liu_atomic_2018}%
  \BibitemOpen
  \bibfield  {author} {\bibinfo {author} {\bibfnamefont {J.}~\bibnamefont
  {Liu}}\ and\ \bibinfo {author} {\bibfnamefont {L.}~\bibnamefont {Cheng}},\
  }\bibfield  {title} {\bibinfo {title} {An atomic mean-field spin-orbit
  approach within exact two-component theory for a non-perturbative treatment
  of spin-orbit coupling},\ }\href {https://doi.org/10.1063/1.5023750}
  {\bibfield  {journal} {\bibinfo  {journal} {J. Chem. Phys.}\ }\textbf
  {\bibinfo {volume} {148}},\ \bibinfo {pages} {144108} (\bibinfo {year}
  {2018})}\BibitemShut {NoStop}%
\bibitem [{\citenamefont {Zhang}\ and\ \citenamefont
  {Cheng}(2022)}]{zhang_atomic_2022}%
  \BibitemOpen
  \bibfield  {author} {\bibinfo {author} {\bibfnamefont {C.}~\bibnamefont
  {Zhang}}\ and\ \bibinfo {author} {\bibfnamefont {L.}~\bibnamefont {Cheng}},\
  }\bibfield  {title} {\bibinfo {title} {Atomic {Mean}-{Field} {Approach}
  within {Exact} {Two}-{Component} {Theory} {Based} on the
  {Dirac}–{Coulomb}–{Breit} {Hamiltonian}},\ }\href
  {https://doi.org/10.1021/acs.jpca.2c02181} {\bibfield  {journal} {\bibinfo
  {journal} {J. Phys. Chem. A}\ }\textbf {\bibinfo {volume} {126}},\ \bibinfo
  {pages} {4537} (\bibinfo {year} {2022})}\BibitemShut {NoStop}%
\bibitem [{\citenamefont {Zhang}\ \emph {et~al.}(2024)\citenamefont {Zhang},
  \citenamefont {Peterson}, \citenamefont {Dyall},\ and\ \citenamefont
  {Cheng}}]{zhang_new_2024}%
  \BibitemOpen
  \bibfield  {author} {\bibinfo {author} {\bibfnamefont {C.}~\bibnamefont
  {Zhang}}, \bibinfo {author} {\bibfnamefont {K.~A.}\ \bibnamefont {Peterson}},
  \bibinfo {author} {\bibfnamefont {K.~G.}\ \bibnamefont {Dyall}},\ and\
  \bibinfo {author} {\bibfnamefont {L.}~\bibnamefont {Cheng}},\ }\href
  {https://doi.org/10.48550/arXiv.2405.04008} {\bibinfo {title} {A new
  computational framework for spinor-based relativistic exact two-component
  calculations using contracted basis functions}} (\bibinfo {year} {2024}),\
  \bibinfo {note} {arXiv:2405.04008 [physics]}\BibitemShut {NoStop}%
\bibitem [{\citenamefont {Dyall}(2004)}]{Dyall04}%
  \BibitemOpen
  \bibfield  {author} {\bibinfo {author} {\bibfnamefont {K.~G.}\ \bibnamefont
  {Dyall}},\ }\bibfield  {title} {\bibinfo {title} {{Relativistic double-zeta,
  triple-zeta, and quadruple-zeta basis sets for the 5d elements Hf–Hg}},\
  }\href {https://doi.org/10.1007/s00214-004-0607-y} {\bibfield  {journal}
  {\bibinfo  {journal} {Theor. Chem. Acc.}\ }\textbf {\bibinfo {volume}
  {112}},\ \bibinfo {pages} {403} (\bibinfo {year} {2004})}\BibitemShut
  {NoStop}%
\bibitem [{\citenamefont {Dyall}(2006)}]{Dyall06}%
  \BibitemOpen
  \bibfield  {author} {\bibinfo {author} {\bibfnamefont {K.~G.}\ \bibnamefont
  {Dyall}},\ }\bibfield  {title} {\bibinfo {title} {{Relativistic
  quadruple-zeta and revised triple-zeta and double-zeta basis sets for the 4p,
  5p, and 6p elements}},\ }\href {https://doi.org/10.1007/s00214-006-0126-0}
  {\bibfield  {journal} {\bibinfo  {journal} {Theor. Chem. Acc.}\ }\textbf
  {\bibinfo {volume} {115}},\ \bibinfo {pages} {441} (\bibinfo {year}
  {2006})}\BibitemShut {NoStop}%
\bibitem [{\citenamefont {{Dunning, Jr.}}(1989)}]{Dunning89}%
  \BibitemOpen
  \bibfield  {author} {\bibinfo {author} {\bibfnamefont {T.~H.}\ \bibnamefont
  {{Dunning, Jr.}}},\ }\bibfield  {title} {\bibinfo {title} {Gaussian basis
  sets for use in correlated molecular calculations. {I. The} atoms boron
  through neon and hydrogen},\ }\href@noop {} {\bibfield  {journal} {\bibinfo
  {journal} {J. Chem. Phys.}\ }\textbf {\bibinfo {volume} {90}},\ \bibinfo
  {pages} {1007} (\bibinfo {year} {1989})}\BibitemShut {NoStop}%
\bibitem [{\citenamefont {Peterson}\ and\ \citenamefont
  {Dunning}(2002)}]{Peterson02}%
  \BibitemOpen
  \bibfield  {author} {\bibinfo {author} {\bibfnamefont {K.~A.}\ \bibnamefont
  {Peterson}}\ and\ \bibinfo {author} {\bibfnamefont {T.~H.}\ \bibnamefont
  {Dunning}},\ }\bibfield  {title} {\bibinfo {title} {{Accurate correlation
  consistent basis sets for molecular core-valence correlation effects: The
  second row atoms Al-Ar, and the first row atoms B-Ne revisited}},\ }\href
  {https://doi.org/10.1063/1.1520138} {\bibfield  {journal} {\bibinfo
  {journal} {J. Chem. Phys.}\ }\textbf {\bibinfo {volume} {117}},\ \bibinfo
  {pages} {10548} (\bibinfo {year} {2002})}\BibitemShut {NoStop}%
\bibitem [{\citenamefont {Dyall}(2001)}]{Dyall01}%
  \BibitemOpen
  \bibfield  {author} {\bibinfo {author} {\bibfnamefont {K.~G.}\ \bibnamefont
  {Dyall}},\ }\bibfield  {title} {\bibinfo {title} {{Interfacing relativistic
  and nonrelativistic methods. IV. One- and two-electron scalar
  approximations}},\ }\href@noop {} {\bibfield  {journal} {\bibinfo  {journal}
  {J. Chem. Phys.}\ }\textbf {\bibinfo {volume} {115}},\ \bibinfo {pages}
  {9136} (\bibinfo {year} {2001})}\BibitemShut {NoStop}%
\bibitem [{\citenamefont {F{\ae}gri}(2001)}]{Faegri01}%
  \BibitemOpen
  \bibfield  {author} {\bibinfo {author} {\bibfnamefont {K.}~\bibnamefont
  {F{\ae}gri}},\ }\bibfield  {title} {\bibinfo {title} {{Relativistic Gaussian
  basis sets for the elements K - Uuo}},\ }\href@noop {} {\bibfield  {journal}
  {\bibinfo  {journal} {Theor. Chem. Acc.}\ }\textbf {\bibinfo {volume}
  {105}},\ \bibinfo {pages} {252} (\bibinfo {year} {2001})}\BibitemShut
  {NoStop}%
\bibitem [{\citenamefont {Roos}\ \emph {et~al.}(2004)\citenamefont {Roos},
  \citenamefont {Lindh}, \citenamefont {Malmqvist}, \citenamefont {Veryazov},\
  and\ \citenamefont {Widmark}}]{Roos04}%
  \BibitemOpen
  \bibfield  {author} {\bibinfo {author} {\bibfnamefont {B.~O.}\ \bibnamefont
  {Roos}}, \bibinfo {author} {\bibfnamefont {R.}~\bibnamefont {Lindh}},
  \bibinfo {author} {\bibfnamefont {P.-{\AA}.}\ \bibnamefont {Malmqvist}},
  \bibinfo {author} {\bibfnamefont {V.}~\bibnamefont {Veryazov}},\ and\
  \bibinfo {author} {\bibfnamefont {P.-O.}\ \bibnamefont {Widmark}},\
  }\bibfield  {title} {\bibinfo {title} {Main group atoms and dimers studied
  with a new relativistic ano basis set},\ }\href@noop {} {\bibfield  {journal}
  {\bibinfo  {journal} {J. Phys. Chem. A}\ }\textbf {\bibinfo {volume} {108}},\
  \bibinfo {pages} {2851} (\bibinfo {year} {2004})}\BibitemShut {NoStop}%
\bibitem [{\citenamefont {Roos}\ \emph {et~al.}(2005)\citenamefont {Roos},
  \citenamefont {Lindh}, \citenamefont {Malmqvist}, \citenamefont {Veryazov},\
  and\ \citenamefont {Widmark}}]{Roos05}%
  \BibitemOpen
  \bibfield  {author} {\bibinfo {author} {\bibfnamefont {B.~O.}\ \bibnamefont
  {Roos}}, \bibinfo {author} {\bibfnamefont {R.}~\bibnamefont {Lindh}},
  \bibinfo {author} {\bibfnamefont {P.-{\AA}.}\ \bibnamefont {Malmqvist}},
  \bibinfo {author} {\bibfnamefont {V.}~\bibnamefont {Veryazov}},\ and\
  \bibinfo {author} {\bibfnamefont {P.-O.}\ \bibnamefont {Widmark}},\
  }\bibfield  {title} {\bibinfo {title} {New relativistic {ANO} basis sets for
  transition metal atoms},\ }\href@noop {} {\bibfield  {journal} {\bibinfo
  {journal} {J. Phys. Chem. A}\ }\textbf {\bibinfo {volume} {109}},\ \bibinfo
  {pages} {6575} (\bibinfo {year} {2005})}\BibitemShut {NoStop}%
\bibitem [{\citenamefont {Liu}\ \emph {et~al.}(2021)\citenamefont {Liu},
  \citenamefont {Zheng}, \citenamefont {Asthana}, \citenamefont {Zhang},\ and\
  \citenamefont {Cheng}}]{liu21}%
  \BibitemOpen
  \bibfield  {author} {\bibinfo {author} {\bibfnamefont {J.}~\bibnamefont
  {Liu}}, \bibinfo {author} {\bibfnamefont {X.}~\bibnamefont {Zheng}}, \bibinfo
  {author} {\bibfnamefont {A.}~\bibnamefont {Asthana}}, \bibinfo {author}
  {\bibfnamefont {C.}~\bibnamefont {Zhang}},\ and\ \bibinfo {author}
  {\bibfnamefont {L.}~\bibnamefont {Cheng}},\ }\bibfield  {title} {\bibinfo
  {title} {Analytic evaluation of energy first derivatives for spin--orbit
  coupled-cluster singles and doubles augmented with noniterative triples
  method: General formulation and an implementation for first-order
  properties},\ }\href@noop {} {\bibfield  {journal} {\bibinfo  {journal} {J.
  Chem. Phys.}\ }\textbf {\bibinfo {volume} {154}},\ \bibinfo {pages} {064110}
  (\bibinfo {year} {2021})}\BibitemShut {NoStop}%
\bibitem [{\citenamefont {Zhang}\ \emph
  {et~al.}(2021{\natexlab{a}})\citenamefont {Zhang}, \citenamefont {Zheng},\
  and\ \citenamefont {Cheng}}]{zhang_calculations_2021}%
  \BibitemOpen
  \bibfield  {author} {\bibinfo {author} {\bibfnamefont {C.}~\bibnamefont
  {Zhang}}, \bibinfo {author} {\bibfnamefont {X.}~\bibnamefont {Zheng}},\ and\
  \bibinfo {author} {\bibfnamefont {L.}~\bibnamefont {Cheng}},\ }\bibfield
  {title} {\bibinfo {title} {Calculations of time-reversal-symmetry-violation
  sensitivity parameters based on analytic relativistic coupled-cluster
  gradient theory},\ }\href {https://doi.org/10.1103/PhysRevA.104.012814}
  {\bibfield  {journal} {\bibinfo  {journal} {Phys. Rev. A}\ }\textbf {\bibinfo
  {volume} {104}},\ \bibinfo {pages} {012814} (\bibinfo {year}
  {2021}{\natexlab{a}})}\BibitemShut {NoStop}%
\bibitem [{\citenamefont {Zhang}\ \emph {et~al.}(2023)\citenamefont {Zhang},
  \citenamefont {Zheng}, \citenamefont {Liu}, \citenamefont {Asthana},\ and\
  \citenamefont {Cheng}}]{Zhang23}%
  \BibitemOpen
  \bibfield  {author} {\bibinfo {author} {\bibfnamefont {C.}~\bibnamefont
  {Zhang}}, \bibinfo {author} {\bibfnamefont {X.}~\bibnamefont {Zheng}},
  \bibinfo {author} {\bibfnamefont {J.}~\bibnamefont {Liu}}, \bibinfo {author}
  {\bibfnamefont {A.}~\bibnamefont {Asthana}},\ and\ \bibinfo {author}
  {\bibfnamefont {L.}~\bibnamefont {Cheng}},\ }\bibfield  {title} {\bibinfo
  {title} {{Analytic gradients for relativistic exact-two-component
  equation-of-motion coupled-cluster singles and doubles method}},\ }\href
  {https://doi.org/10.1063/5.0175041} {\bibfield  {journal} {\bibinfo
  {journal} {J. Chem. Phys.}\ }\textbf {\bibinfo {volume} {159}},\ \bibinfo
  {pages} {244113} (\bibinfo {year} {2023})}\BibitemShut {NoStop}%
\bibitem [{\citenamefont {Stanton}\ and\ \citenamefont
  {Bartlett}(1993)}]{Stanton93a}%
  \BibitemOpen
  \bibfield  {author} {\bibinfo {author} {\bibfnamefont {J.~F.}\ \bibnamefont
  {Stanton}}\ and\ \bibinfo {author} {\bibfnamefont {R.~J.}\ \bibnamefont
  {Bartlett}},\ }\bibfield  {title} {\bibinfo {title} {The equation of motion
  coupled-cluster method. {A} systematic biorthogonal approach to molecular
  excitation energies, transition probabilities, and excited state
  properties},\ }\href@noop {} {\bibfield  {journal} {\bibinfo  {journal} {J.
  Chem. Phys.}\ }\textbf {\bibinfo {volume} {98}},\ \bibinfo {pages} {7029}
  (\bibinfo {year} {1993})}\BibitemShut {NoStop}%
\bibitem [{\citenamefont {Nooijen}\ and\ \citenamefont
  {Bartlett}(1995)}]{Nooijen95}%
  \BibitemOpen
  \bibfield  {author} {\bibinfo {author} {\bibfnamefont {M.}~\bibnamefont
  {Nooijen}}\ and\ \bibinfo {author} {\bibfnamefont {R.~J.}\ \bibnamefont
  {Bartlett}},\ }\bibfield  {title} {\bibinfo {title} {Equation of motion
  coupled cluster method for electron attachment},\ }\href@noop {} {\bibfield
  {journal} {\bibinfo  {journal} {J. Chem. Phys.}\ }\textbf {\bibinfo {volume}
  {102}},\ \bibinfo {pages} {3629} (\bibinfo {year} {1995})}\BibitemShut
  {NoStop}%
\bibitem [{\citenamefont {Stanton}\ and\ \citenamefont
  {Gauss}(1999)}]{Stanton99}%
  \BibitemOpen
  \bibfield  {author} {\bibinfo {author} {\bibfnamefont {J.~F.}\ \bibnamefont
  {Stanton}}\ and\ \bibinfo {author} {\bibfnamefont {J.}~\bibnamefont
  {Gauss}},\ }\bibfield  {title} {\bibinfo {title} {A simple scheme for the
  direct calculation of ionization potentials with coupled-cluster theory that
  exploits established excitation energy methods},\ }\href@noop {} {\bibfield
  {journal} {\bibinfo  {journal} {J. Chem. Phys.}\ }\textbf {\bibinfo {volume}
  {111}},\ \bibinfo {pages} {8785} (\bibinfo {year} {1999})}\BibitemShut
  {NoStop}%
\bibitem [{\citenamefont {Asthana}\ \emph {et~al.}(2019)\citenamefont
  {Asthana}, \citenamefont {Liu},\ and\ \citenamefont {Cheng}}]{Asthana19}%
  \BibitemOpen
  \bibfield  {author} {\bibinfo {author} {\bibfnamefont {A.}~\bibnamefont
  {Asthana}}, \bibinfo {author} {\bibfnamefont {J.}~\bibnamefont {Liu}},\ and\
  \bibinfo {author} {\bibfnamefont {L.}~\bibnamefont {Cheng}},\ }\bibfield
  {title} {\bibinfo {title} {Exact two-component equation-of-motion
  coupled-cluster singles and doubles method using atomic mean-field spin-orbit
  integrals},\ }\href@noop {} {\bibfield  {journal} {\bibinfo  {journal} {J.
  Chem. Phys.}\ }\textbf {\bibinfo {volume} {150}},\ \bibinfo {pages} {074102}
  (\bibinfo {year} {2019})}\BibitemShut {NoStop}%
\bibitem [{\citenamefont {Zhang}\ \emph {et~al.}(2020)\citenamefont {Zhang},
  \citenamefont {Korslund}, \citenamefont {Wu}, \citenamefont {Ding},\ and\
  \citenamefont {Cheng}}]{zhang_towards_2020}%
  \BibitemOpen
  \bibfield  {author} {\bibinfo {author} {\bibfnamefont {C.}~\bibnamefont
  {Zhang}}, \bibinfo {author} {\bibfnamefont {H.}~\bibnamefont {Korslund}},
  \bibinfo {author} {\bibfnamefont {Y.}~\bibnamefont {Wu}}, \bibinfo {author}
  {\bibfnamefont {S.}~\bibnamefont {Ding}},\ and\ \bibinfo {author}
  {\bibfnamefont {L.}~\bibnamefont {Cheng}},\ }\bibfield  {title} {\bibinfo
  {title} {Towards accurate prediction for laser-coolable molecules:
  relativistic coupled-cluster calculations for yttrium monoxide and prospects
  for improving its laser cooling efficiencies},\ }\href
  {https://doi.org/10.1039/D0CP04608F} {\bibfield  {journal} {\bibinfo
  {journal} {Phys. Chem. Chem. Phys.}\ }\textbf {\bibinfo {volume} {22}},\
  \bibinfo {pages} {26167} (\bibinfo {year} {2020})}\BibitemShut {NoStop}%
\bibitem [{\citenamefont {Zhang}\ \emph
  {et~al.}(2021{\natexlab{b}})\citenamefont {Zhang}, \citenamefont
  {Augenbraun}, \citenamefont {Lasner}, \citenamefont {Vilas}, \citenamefont
  {Doyle},\ and\ \citenamefont {Cheng}}]{zhang_accurate_2021}%
  \BibitemOpen
  \bibfield  {author} {\bibinfo {author} {\bibfnamefont {C.}~\bibnamefont
  {Zhang}}, \bibinfo {author} {\bibfnamefont {B.~L.}\ \bibnamefont
  {Augenbraun}}, \bibinfo {author} {\bibfnamefont {Z.~D.}\ \bibnamefont
  {Lasner}}, \bibinfo {author} {\bibfnamefont {N.~B.}\ \bibnamefont {Vilas}},
  \bibinfo {author} {\bibfnamefont {J.~M.}\ \bibnamefont {Doyle}},\ and\
  \bibinfo {author} {\bibfnamefont {L.}~\bibnamefont {Cheng}},\ }\bibfield
  {title} {\bibinfo {title} {Accurate prediction and measurement of vibronic
  branching ratios for laser cooling linear polyatomic molecules},\ }\href
  {https://doi.org/10.1063/5.0063611} {\bibfield  {journal} {\bibinfo
  {journal} {J. Chem. Phys.}\ }\textbf {\bibinfo {volume} {155}},\ \bibinfo
  {pages} {091101} (\bibinfo {year} {2021}{\natexlab{b}})}\BibitemShut
  {NoStop}%
\bibitem [{\citenamefont {Mengesha}\ \emph {et~al.}(2020)\citenamefont
  {Mengesha}, \citenamefont {Le}, \citenamefont {Steimle}, \citenamefont
  {Cheng}, \citenamefont {Zhang}, \citenamefont {Augenbraun}, \citenamefont
  {Lasner},\ and\ \citenamefont {Doyle}}]{mengesha2020branching}%
  \BibitemOpen
  \bibfield  {author} {\bibinfo {author} {\bibfnamefont {E.~T.}\ \bibnamefont
  {Mengesha}}, \bibinfo {author} {\bibfnamefont {A.~T.}\ \bibnamefont {Le}},
  \bibinfo {author} {\bibfnamefont {T.~C.}\ \bibnamefont {Steimle}}, \bibinfo
  {author} {\bibfnamefont {L.}~\bibnamefont {Cheng}}, \bibinfo {author}
  {\bibfnamefont {C.}~\bibnamefont {Zhang}}, \bibinfo {author} {\bibfnamefont
  {B.~L.}\ \bibnamefont {Augenbraun}}, \bibinfo {author} {\bibfnamefont
  {Z.}~\bibnamefont {Lasner}},\ and\ \bibinfo {author} {\bibfnamefont
  {J.}~\bibnamefont {Doyle}},\ }\bibfield  {title} {\bibinfo {title} {Branching
  {Ratios}, {Radiative} {Lifetimes}, and {Transition} {Dipole} {Moments} for
  {YbOH}},\ }\href {https://doi.org/10.1021/acs.jpca.0c00850} {\bibfield
  {journal} {\bibinfo  {journal} {J. Phys. Chem. A}\ }\textbf {\bibinfo
  {volume} {124}},\ \bibinfo {pages} {3135} (\bibinfo {year}
  {2020})}\BibitemShut {NoStop}%
\bibitem [{\citenamefont {{Purvis III}}\ and\ \citenamefont
  {J.Bartlett}(1982)}]{Purvis82}%
  \BibitemOpen
  \bibfield  {author} {\bibinfo {author} {\bibfnamefont {G.~D.}\ \bibnamefont
  {{Purvis III}}}\ and\ \bibinfo {author} {\bibfnamefont {R.}~\bibnamefont
  {J.Bartlett}},\ }\bibfield  {title} {\bibinfo {title} {A full
  coupled--cluster singles and doubles model: {The} inclusion of disconnected
  triples},\ }\href@noop {} {\bibfield  {journal} {\bibinfo  {journal} {J.
  Chem. Phys.}\ }\textbf {\bibinfo {volume} {76}},\ \bibinfo {pages} {1910}
  (\bibinfo {year} {1982})}\BibitemShut {NoStop}%
\bibitem [{\citenamefont {Raghavachari}\ \emph {et~al.}(1989)\citenamefont
  {Raghavachari}, \citenamefont {Trucks}, \citenamefont {Pople},\ and\
  \citenamefont {Head-Gordon}}]{Raghavachari89}%
  \BibitemOpen
  \bibfield  {author} {\bibinfo {author} {\bibfnamefont {K.}~\bibnamefont
  {Raghavachari}}, \bibinfo {author} {\bibfnamefont {G.~W.}\ \bibnamefont
  {Trucks}}, \bibinfo {author} {\bibfnamefont {J.~A.}\ \bibnamefont {Pople}},\
  and\ \bibinfo {author} {\bibfnamefont {M.}~\bibnamefont {Head-Gordon}},\
  }\bibfield  {title} {\bibinfo {title} {A fifth-order perturbation comparison
  of electron correlation thoeries},\ }\href
  {https://doi.org/10.1016/S0009-2614(89)87395-6} {\bibfield  {journal}
  {\bibinfo  {journal} {Chem. Phys. Lett.}\ }\textbf {\bibinfo {volume}
  {157}},\ \bibinfo {pages} {479} (\bibinfo {year} {1989})}\BibitemShut
  {NoStop}%
\bibitem [{\citenamefont {Peterson}\ and\ \citenamefont
  {Puzzarini}(2005)}]{Peterson2005}%
  \BibitemOpen
  \bibfield  {author} {\bibinfo {author} {\bibfnamefont {K.~A.}\ \bibnamefont
  {Peterson}}\ and\ \bibinfo {author} {\bibfnamefont {C.}~\bibnamefont
  {Puzzarini}},\ }\bibfield  {title} {\bibinfo {title} {{Systematically
  convergent basis sets for transition metals. II. Pseudopotential-based
  correlation consistent basis sets for the group 11 (Cu, Ag, Au) and 12 (Zn,
  Cd, Hg) elements}},\ }\href {https://doi.org/10.1007/s00214-005-0681-9}
  {\bibfield  {journal} {\bibinfo  {journal} {Theor. Chem. Acc.}\ }\textbf
  {\bibinfo {volume} {114}},\ \bibinfo {pages} {283–296} (\bibinfo {year}
  {2005})}\BibitemShut {NoStop}%
\bibitem [{\citenamefont {Houdart}\ and\ \citenamefont
  {Schamps}(1973)}]{houdart_emission_1973}%
  \BibitemOpen
  \bibfield  {author} {\bibinfo {author} {\bibfnamefont {R.}~\bibnamefont
  {Houdart}}\ and\ \bibinfo {author} {\bibfnamefont {J.}~\bibnamefont
  {Schamps}},\ }\bibfield  {title} {\bibinfo {title} {The emission spectra of
  the {AuSi}, {AuGe}, {AuSn} and {AuPb} radicals},\ }\href
  {https://doi.org/10.1088/0022-3700/6/11/045} {\bibfield  {journal} {\bibinfo
  {journal} {J. Phys. B: At. Mol. Phys.}\ }\textbf {\bibinfo {volume} {6}},\
  \bibinfo {pages} {2478} (\bibinfo {year} {1973})}\BibitemShut {NoStop}%
\bibitem [{\citenamefont {Tannor}(2008)}]{tannor_introduction_2008}%
  \BibitemOpen
  \bibfield  {author} {\bibinfo {author} {\bibfnamefont {D.}~\bibnamefont
  {Tannor}},\ }\href@noop {} {\emph {\bibinfo {title} {Introduction to
  {Quantum} {Mechanics}: a {Time}-{Dependent} {Perspective}}}}\ (\bibinfo
  {publisher} {University Science Books},\ \bibinfo {year} {2008})\ \bibinfo
  {note} {oCLC: 1413373826}\BibitemShut {NoStop}%
\bibitem [{\citenamefont {Barrow}\ \emph {et~al.}(1964)\citenamefont {Barrow},
  \citenamefont {Gissane},\ and\ \citenamefont
  {Travis}}]{barrow_electronic_1964}%
  \BibitemOpen
  \bibfield  {author} {\bibinfo {author} {\bibfnamefont {R.~F.}\ \bibnamefont
  {Barrow}}, \bibinfo {author} {\bibfnamefont {W.~J.~M.}\ \bibnamefont
  {Gissane}},\ and\ \bibinfo {author} {\bibfnamefont {D.~N.}\ \bibnamefont
  {Travis}},\ }\bibfield  {title} {\bibinfo {title} {Electronic {Spectra} of
  some {Gaseous} {Diatomic} {Compounds} of {Gold}},\ }\href
  {https://doi.org/10.1038/201603a0} {\bibfield  {journal} {\bibinfo  {journal}
  {Nature}\ }\textbf {\bibinfo {volume} {201}},\ \bibinfo {pages} {603}
  (\bibinfo {year} {1964})}\BibitemShut {NoStop}%
\bibitem [{\citenamefont {Barysz}\ \emph {et~al.}(2021)\citenamefont {Barysz},
  \citenamefont {Cernusak}, \citenamefont {Kello},\ and\ \citenamefont
  {Neogrady}}]{barysz_ausi_2021}%
  \BibitemOpen
  \bibfield  {author} {\bibinfo {author} {\bibfnamefont {M.}~\bibnamefont
  {Barysz}}, \bibinfo {author} {\bibfnamefont {I.}~\bibnamefont {Cernusak}},
  \bibinfo {author} {\bibfnamefont {V.}~\bibnamefont {Kello}},\ and\ \bibinfo
  {author} {\bibfnamefont {P.}~\bibnamefont {Neogrady}},\ }\bibfield  {title}
  {\bibinfo {title} {{AuSi} molecule revisited: {IOTC}/{CASSCF}/{CASPT2}
  calculations},\ }\href {https://doi.org/10.1002/qua.26502} {\bibfield
  {journal} {\bibinfo  {journal} {Int. J. Quantum Chem}\ }\textbf {\bibinfo
  {volume} {121}},\ \bibinfo {pages} {e26502} (\bibinfo {year}
  {2021})}\BibitemShut {NoStop}%
\bibitem [{\citenamefont {Tran}\ \emph {et~al.}(2018)\citenamefont {Tran},
  \citenamefont {Lu}, \citenamefont {Zhao}, \citenamefont {Xu}, \citenamefont
  {Xu}, \citenamefont {Tran}, \citenamefont {Li},\ and\ \citenamefont
  {Zheng}}]{tran_spinorbit_2018}%
  \BibitemOpen
  \bibfield  {author} {\bibinfo {author} {\bibfnamefont {Q.~T.}\ \bibnamefont
  {Tran}}, \bibinfo {author} {\bibfnamefont {S.-J.}\ \bibnamefont {Lu}},
  \bibinfo {author} {\bibfnamefont {L.-J.}\ \bibnamefont {Zhao}}, \bibinfo
  {author} {\bibfnamefont {X.-L.}\ \bibnamefont {Xu}}, \bibinfo {author}
  {\bibfnamefont {H.-G.}\ \bibnamefont {Xu}}, \bibinfo {author} {\bibfnamefont
  {V.~T.}\ \bibnamefont {Tran}}, \bibinfo {author} {\bibfnamefont
  {J.}~\bibnamefont {Li}},\ and\ \bibinfo {author} {\bibfnamefont {W.-J.}\
  \bibnamefont {Zheng}},\ }\bibfield  {title} {\bibinfo {title} {Spin–{Orbit}
  {Splittings} and {Low}-{Lying} {Electronic} {States} of {AuSi} and {AuGe}:
  {Anion} {Photoelectron} {Spectroscopy} and \textit{ab {Initio}}
  {Calculations}},\ }\href {https://doi.org/10.1021/acs.jpca.8b01366}
  {\bibfield  {journal} {\bibinfo  {journal} {J. Phys. Chem. A}\ }\textbf
  {\bibinfo {volume} {122}},\ \bibinfo {pages} {3374} (\bibinfo {year}
  {2018})}\BibitemShut {NoStop}%
\bibitem [{\citenamefont {Scherer}\ \emph {et~al.}(1995)\citenamefont
  {Scherer}, \citenamefont {Paul}, \citenamefont {Collier}, \citenamefont
  {O’Keefe},\ and\ \citenamefont {Saykally}}]{scherer_cavity_1995}%
  \BibitemOpen
  \bibfield  {author} {\bibinfo {author} {\bibfnamefont {J.~J.}\ \bibnamefont
  {Scherer}}, \bibinfo {author} {\bibfnamefont {J.~B.}\ \bibnamefont {Paul}},
  \bibinfo {author} {\bibfnamefont {C.~P.}\ \bibnamefont {Collier}}, \bibinfo
  {author} {\bibfnamefont {A.}~\bibnamefont {O’Keefe}},\ and\ \bibinfo
  {author} {\bibfnamefont {R.~J.}\ \bibnamefont {Saykally}},\ }\bibfield
  {title} {\bibinfo {title} {Cavity ringdown laser absorption spectroscopy and
  time‐of‐flight mass spectroscopy of jet‐cooled gold silicides},\ }\href
  {https://doi.org/10.1063/1.470029} {\bibfield  {journal} {\bibinfo  {journal}
  {J. Chem. Phys.}\ }\textbf {\bibinfo {volume} {103}},\ \bibinfo {pages}
  {9187} (\bibinfo {year} {1995})}\BibitemShut {NoStop}%
\bibitem [{\citenamefont {Boldyrev}\ \emph {et~al.}(1998)\citenamefont
  {Boldyrev}, \citenamefont {Simons}, \citenamefont {Scherer}, \citenamefont
  {Paul}, \citenamefont {Collier},\ and\ \citenamefont
  {Saykally}}]{boldyrev_ground_1998}%
  \BibitemOpen
  \bibfield  {author} {\bibinfo {author} {\bibfnamefont {A.~I.}\ \bibnamefont
  {Boldyrev}}, \bibinfo {author} {\bibfnamefont {J.}~\bibnamefont {Simons}},
  \bibinfo {author} {\bibfnamefont {J.~J.}\ \bibnamefont {Scherer}}, \bibinfo
  {author} {\bibfnamefont {J.~B.}\ \bibnamefont {Paul}}, \bibinfo {author}
  {\bibfnamefont {C.~P.}\ \bibnamefont {Collier}},\ and\ \bibinfo {author}
  {\bibfnamefont {R.~J.}\ \bibnamefont {Saykally}},\ }\bibfield  {title}
  {\bibinfo {title} {On the ground electronic states of copper silicide and its
  ions},\ }\href {https://doi.org/10.1063/1.475982} {\bibfield  {journal}
  {\bibinfo  {journal} {J. Chem. Phys.}\ }\textbf {\bibinfo {volume} {108}},\
  \bibinfo {pages} {5728} (\bibinfo {year} {1998})}\BibitemShut {NoStop}%
\bibitem [{\citenamefont {Li}\ \emph {et~al.}(2011)\citenamefont {Li},
  \citenamefont {Zhang}, \citenamefont {Meng},\ and\ \citenamefont
  {Yu}}]{li_electronic_2011}%
  \BibitemOpen
  \bibfield  {author} {\bibinfo {author} {\bibfnamefont {Z.}~\bibnamefont
  {Li}}, \bibinfo {author} {\bibfnamefont {J.}~\bibnamefont {Zhang}}, \bibinfo
  {author} {\bibfnamefont {D.}~\bibnamefont {Meng}},\ and\ \bibinfo {author}
  {\bibfnamefont {Y.}~\bibnamefont {Yu}},\ }\bibfield  {title} {\bibinfo
  {title} {Electronic structure and bonding characters of the two lowest states
  of copper, silver, and gold monocarbides},\ }\href
  {https://doi.org/10.1016/j.comptc.2011.02.019} {\bibfield  {journal}
  {\bibinfo  {journal} {Computational and Theoretical Chemistry}\ }\textbf
  {\bibinfo {volume} {966}},\ \bibinfo {pages} {97} (\bibinfo {year}
  {2011})}\BibitemShut {NoStop}%
\bibitem [{noa(2015)}]{noauthor_atomic_2015}%
  \BibitemOpen
  \bibfield  {title} {\bibinfo {title} {Atomic {Reference} {Data} for
  {Electronic} {Structure} {Calculations}, {Atomic} {Total} {Energies} and
  {Eigenvalues} ({HTML})},\ }\href
  {https://www.nist.gov/pml/atomic-reference-data-electronic-structure-calculations/atomic-reference-data-electronic-7}
  {\bibfield  {journal} {\bibinfo  {journal} {NIST}\ } (\bibinfo {year}
  {2015})},\ \bibinfo {note} {last Modified:
  2017-04-24T15:25-04:00}\BibitemShut {NoStop}%
\bibitem [{\citenamefont {Wu}\ and\ \citenamefont
  {Su}(2006)}]{wu_electronic_2006}%
  \BibitemOpen
  \bibfield  {author} {\bibinfo {author} {\bibfnamefont {Z.~J.}\ \bibnamefont
  {Wu}}\ and\ \bibinfo {author} {\bibfnamefont {Z.~M.}\ \bibnamefont {Su}},\
  }\bibfield  {title} {\bibinfo {title} {Electronic structures and chemical
  bonding in transition metal monosilicides {MSi} ({M}=3d, 4d, 5d elements)},\
  }\href {https://doi.org/10.1063/1.2196040} {\bibfield  {journal} {\bibinfo
  {journal} {J. Chem. Phys.}\ }\textbf {\bibinfo {volume} {124}},\ \bibinfo
  {pages} {184306} (\bibinfo {year} {2006})}\BibitemShut {NoStop}%
\bibitem [{\citenamefont {Abe}\ \emph {et~al.}(2002)\citenamefont {Abe},
  \citenamefont {Nakajima},\ and\ \citenamefont
  {Hirao}}]{abe_theoretical_2002}%
  \BibitemOpen
  \bibfield  {author} {\bibinfo {author} {\bibfnamefont {M.}~\bibnamefont
  {Abe}}, \bibinfo {author} {\bibfnamefont {T.}~\bibnamefont {Nakajima}},\ and\
  \bibinfo {author} {\bibfnamefont {K.}~\bibnamefont {Hirao}},\ }\bibfield
  {title} {\bibinfo {title} {A theoretical study of the low-lying states of the
  {AuSi} molecule: {An} assignment of the excited \textit{{A}} and \textit{{D}}
  states},\ }\href {http://aip.scitation.org/doi/10.1063/1.1494981} {\bibfield
  {journal} {\bibinfo  {journal} {J. Chem. Phys.}\ }\textbf {\bibinfo {volume}
  {117}},\ \bibinfo {pages} {7960} (\bibinfo {year} {2002})}\BibitemShut
  {NoStop}%
\bibitem [{\citenamefont {Barry}\ \emph {et~al.}(2014)\citenamefont {Barry},
  \citenamefont {McCarron}, \citenamefont {Norrgard}, \citenamefont
  {Steinecker},\ and\ \citenamefont {DeMille}}]{barry2014magnetooptical}%
  \BibitemOpen
  \bibfield  {author} {\bibinfo {author} {\bibfnamefont {J.~F.}\ \bibnamefont
  {Barry}}, \bibinfo {author} {\bibfnamefont {D.~J.}\ \bibnamefont {McCarron}},
  \bibinfo {author} {\bibfnamefont {E.~B.}\ \bibnamefont {Norrgard}}, \bibinfo
  {author} {\bibfnamefont {M.~H.}\ \bibnamefont {Steinecker}},\ and\ \bibinfo
  {author} {\bibfnamefont {D.}~\bibnamefont {DeMille}},\ }\bibfield  {title}
  {\bibinfo {title} {Magneto-optical trapping of a diatomic molecule},\ }\href
  {https://doi.org/10.1038/nature13634} {\bibfield  {journal} {\bibinfo
  {journal} {Nature}\ }\textbf {\bibinfo {volume} {512}},\ \bibinfo {pages}
  {286} (\bibinfo {year} {2014})}\BibitemShut {NoStop}%
\bibitem [{\citenamefont {Truppe}\ \emph {et~al.}(2017)\citenamefont {Truppe},
  \citenamefont {Williams}, \citenamefont {Hambach}, \citenamefont {Caldwell},
  \citenamefont {Fitch}, \citenamefont {Hinds}, \citenamefont {Sauer},\ and\
  \citenamefont {Tarbutt}}]{truppe_molecules_2017}%
  \BibitemOpen
  \bibfield  {author} {\bibinfo {author} {\bibfnamefont {S.}~\bibnamefont
  {Truppe}}, \bibinfo {author} {\bibfnamefont {H.~J.}\ \bibnamefont
  {Williams}}, \bibinfo {author} {\bibfnamefont {M.}~\bibnamefont {Hambach}},
  \bibinfo {author} {\bibfnamefont {L.}~\bibnamefont {Caldwell}}, \bibinfo
  {author} {\bibfnamefont {N.~J.}\ \bibnamefont {Fitch}}, \bibinfo {author}
  {\bibfnamefont {E.~A.}\ \bibnamefont {Hinds}}, \bibinfo {author}
  {\bibfnamefont {B.~E.}\ \bibnamefont {Sauer}},\ and\ \bibinfo {author}
  {\bibfnamefont {M.~R.}\ \bibnamefont {Tarbutt}},\ }\bibfield  {title}
  {\bibinfo {title} {Molecules cooled below the {Doppler} limit},\ }\href
  {https://doi.org/10.1038/nphys4241} {\bibfield  {journal} {\bibinfo
  {journal} {Nature Phys}\ }\textbf {\bibinfo {volume} {13}},\ \bibinfo {pages}
  {1173} (\bibinfo {year} {2017})}\BibitemShut {NoStop}%
\bibitem [{\citenamefont {Lim}\ \emph {et~al.}(2018)\citenamefont {Lim},
  \citenamefont {Almond}, \citenamefont {Trigatzis}, \citenamefont {Devlin},
  \citenamefont {Fitch}, \citenamefont {Sauer}, \citenamefont {Tarbutt},\ and\
  \citenamefont {Hinds}}]{lim2018laser}%
  \BibitemOpen
  \bibfield  {author} {\bibinfo {author} {\bibfnamefont {J.}~\bibnamefont
  {Lim}}, \bibinfo {author} {\bibfnamefont {J.~R.}\ \bibnamefont {Almond}},
  \bibinfo {author} {\bibfnamefont {M.~A.}\ \bibnamefont {Trigatzis}}, \bibinfo
  {author} {\bibfnamefont {J.~A.}\ \bibnamefont {Devlin}}, \bibinfo {author}
  {\bibfnamefont {N.~J.}\ \bibnamefont {Fitch}}, \bibinfo {author}
  {\bibfnamefont {B.~E.}\ \bibnamefont {Sauer}}, \bibinfo {author}
  {\bibfnamefont {M.~R.}\ \bibnamefont {Tarbutt}},\ and\ \bibinfo {author}
  {\bibfnamefont {E.~A.}\ \bibnamefont {Hinds}},\ }\bibfield  {title} {\bibinfo
  {title} {Laser cooled {YbF} molecules for measuring the electron's electric
  dipole moment},\ }\href {https://doi.org/10.1103/PhysRevLett.120.123201}
  {\bibfield  {journal} {\bibinfo  {journal} {Phys. Rev. Lett.}\ }\textbf
  {\bibinfo {volume} {120}},\ \bibinfo {pages} {123201} (\bibinfo {year}
  {2018})}\BibitemShut {NoStop}%
\bibitem [{\citenamefont {Collopy}\ \emph {et~al.}(2018)\citenamefont
  {Collopy}, \citenamefont {Ding}, \citenamefont {Wu}, \citenamefont
  {Finneran}, \citenamefont {Anderegg}, \citenamefont {Augenbraun},
  \citenamefont {Doyle},\ and\ \citenamefont {Ye}}]{collopy20183d}%
  \BibitemOpen
  \bibfield  {author} {\bibinfo {author} {\bibfnamefont {A.~L.}\ \bibnamefont
  {Collopy}}, \bibinfo {author} {\bibfnamefont {S.}~\bibnamefont {Ding}},
  \bibinfo {author} {\bibfnamefont {Y.}~\bibnamefont {Wu}}, \bibinfo {author}
  {\bibfnamefont {I.~A.}\ \bibnamefont {Finneran}}, \bibinfo {author}
  {\bibfnamefont {L.}~\bibnamefont {Anderegg}}, \bibinfo {author}
  {\bibfnamefont {B.~L.}\ \bibnamefont {Augenbraun}}, \bibinfo {author}
  {\bibfnamefont {J.~M.}\ \bibnamefont {Doyle}},\ and\ \bibinfo {author}
  {\bibfnamefont {J.}~\bibnamefont {Ye}},\ }\bibfield  {title} {\bibinfo
  {title} {{3D} {Magneto}-{Optical} {Trap} of {Yttrium} {Monoxide}},\ }\href
  {doi.org/10.1103/PhysRevLett.121.213201} {\bibfield  {journal} {\bibinfo
  {journal} {Phys. Rev. Lett.}\ }\textbf {\bibinfo {volume} {121}},\ \bibinfo
  {pages} {213201} (\bibinfo {year} {2018})}\BibitemShut {NoStop}%
\bibitem [{\citenamefont {Augenbraun}\ \emph
  {et~al.}(2020{\natexlab{a}})\citenamefont {Augenbraun}, \citenamefont
  {Lasner}, \citenamefont {Frenett}, \citenamefont {Sawaoka}, \citenamefont
  {Miller}, \citenamefont {Steimle},\ and\ \citenamefont
  {Doyle}}]{augenbraun2020lasercooled}%
  \BibitemOpen
  \bibfield  {author} {\bibinfo {author} {\bibfnamefont {B.~L.}\ \bibnamefont
  {Augenbraun}}, \bibinfo {author} {\bibfnamefont {Z.~D.}\ \bibnamefont
  {Lasner}}, \bibinfo {author} {\bibfnamefont {A.}~\bibnamefont {Frenett}},
  \bibinfo {author} {\bibfnamefont {H.}~\bibnamefont {Sawaoka}}, \bibinfo
  {author} {\bibfnamefont {C.}~\bibnamefont {Miller}}, \bibinfo {author}
  {\bibfnamefont {T.~C.}\ \bibnamefont {Steimle}},\ and\ \bibinfo {author}
  {\bibfnamefont {J.~M.}\ \bibnamefont {Doyle}},\ }\bibfield  {title} {\bibinfo
  {title} {Laser-cooled polyatomic molecules for improved electron electric
  dipole moment searches},\ }\href
  {https://iopscience.iop.org/article/10.1088/1367-2630/ab687b} {\bibfield
  {journal} {\bibinfo  {journal} {New J. Phys.}\ }\textbf {\bibinfo {volume}
  {22}} (\bibinfo {year} {2020}{\natexlab{a}})}\BibitemShut {NoStop}%
\bibitem [{\citenamefont {Vázquez-Carson}\ \emph {et~al.}(2022)\citenamefont
  {Vázquez-Carson}, \citenamefont {Sun}, \citenamefont {Dai}, \citenamefont
  {Mitra},\ and\ \citenamefont {Zelevinsky}}]{vazquez-carson_direct_2022}%
  \BibitemOpen
  \bibfield  {author} {\bibinfo {author} {\bibfnamefont {S.~F.}\ \bibnamefont
  {Vázquez-Carson}}, \bibinfo {author} {\bibfnamefont {Q.}~\bibnamefont
  {Sun}}, \bibinfo {author} {\bibfnamefont {J.}~\bibnamefont {Dai}}, \bibinfo
  {author} {\bibfnamefont {D.}~\bibnamefont {Mitra}},\ and\ \bibinfo {author}
  {\bibfnamefont {T.}~\bibnamefont {Zelevinsky}},\ }\bibfield  {title}
  {\bibinfo {title} {Direct laser cooling of calcium monohydride molecules},\
  }\href {https://doi.org/10.1088/1367-2630/ac806c} {\bibfield  {journal}
  {\bibinfo  {journal} {New J. Phys.}\ }\textbf {\bibinfo {volume} {24}},\
  \bibinfo {pages} {083006} (\bibinfo {year} {2022})}\BibitemShut {NoStop}%
\bibitem [{\citenamefont
  {Schwerdtfeger}(2002)}]{schwerdtfeger_relativistic_2002}%
  \BibitemOpen
  \bibfield  {author} {\bibinfo {author} {\bibfnamefont {P.}~\bibnamefont
  {Schwerdtfeger}},\ }\bibfield  {title} {\bibinfo {title} {Relativistic
  effects in properties of gold},\ }\href {https://doi.org/10.1002/hc.10093}
  {\bibfield  {journal} {\bibinfo  {journal} {Heteroatom Chemistry}\ }\textbf
  {\bibinfo {volume} {13}},\ \bibinfo {pages} {578} (\bibinfo {year}
  {2002})}\BibitemShut {NoStop}%
\bibitem [{\citenamefont {Bartlett}(1998)}]{bartlett_relativistic_1998}%
  \BibitemOpen
  \bibfield  {author} {\bibinfo {author} {\bibfnamefont {N.}~\bibnamefont
  {Bartlett}},\ }\bibfield  {title} {\bibinfo {title} {Relativistic effects and
  the chemistry of gold},\ }\href {https://doi.org/10.1007/BF03215471}
  {\bibfield  {journal} {\bibinfo  {journal} {Gold Bull}\ }\textbf {\bibinfo
  {volume} {31}},\ \bibinfo {pages} {22} (\bibinfo {year} {1998})}\BibitemShut
  {NoStop}%
\bibitem [{\citenamefont {Pyykkö}(2012)}]{pyykko_relativistic_2012}%
  \BibitemOpen
  \bibfield  {author} {\bibinfo {author} {\bibfnamefont {P.}~\bibnamefont
  {Pyykkö}},\ }\bibfield  {title} {\bibinfo {title} {Relativistic {Effects} in
  {Chemistry}: {More} {Common} {Than} {You} {Thought}},\ }\href
  {https://doi.org/10.1146/annurev-physchem-032511-143755} {\bibfield
  {journal} {\bibinfo  {journal} {Annu. Rev. Phys. Chem.}\ }\textbf {\bibinfo
  {volume} {63}},\ \bibinfo {pages} {45} (\bibinfo {year} {2012})}\BibitemShut
  {NoStop}%
\bibitem [{\citenamefont {Byrnes}\ \emph {et~al.}(1999)\citenamefont {Byrnes},
  \citenamefont {Dzuba}, \citenamefont {Flambaum},\ and\ \citenamefont
  {Murray}}]{byrnes_enhancement_1999}%
  \BibitemOpen
  \bibfield  {author} {\bibinfo {author} {\bibfnamefont {T.~M.~R.}\
  \bibnamefont {Byrnes}}, \bibinfo {author} {\bibfnamefont {V.~A.}\
  \bibnamefont {Dzuba}}, \bibinfo {author} {\bibfnamefont {V.~V.}\ \bibnamefont
  {Flambaum}},\ and\ \bibinfo {author} {\bibfnamefont {D.~W.}\ \bibnamefont
  {Murray}},\ }\bibfield  {title} {\bibinfo {title} {Enhancement factor for the
  electron electric dipole moment in francium and gold atoms},\ }\href
  {https://doi.org/10.1103/PhysRevA.59.3082} {\bibfield  {journal} {\bibinfo
  {journal} {Phys. Rev. A}\ }\textbf {\bibinfo {volume} {59}},\ \bibinfo
  {pages} {3082} (\bibinfo {year} {1999})}\BibitemShut {NoStop}%
\bibitem [{\citenamefont {Hudson}\ \emph {et~al.}(2011)\citenamefont {Hudson},
  \citenamefont {Kara}, \citenamefont {Smallman}, \citenamefont {Sauer},
  \citenamefont {Tarbutt},\ and\ \citenamefont {Hinds}}]{hudson_improved_2011}%
  \BibitemOpen
  \bibfield  {author} {\bibinfo {author} {\bibfnamefont {J.~J.}\ \bibnamefont
  {Hudson}}, \bibinfo {author} {\bibfnamefont {D.~M.}\ \bibnamefont {Kara}},
  \bibinfo {author} {\bibfnamefont {I.~J.}\ \bibnamefont {Smallman}}, \bibinfo
  {author} {\bibfnamefont {B.~E.}\ \bibnamefont {Sauer}}, \bibinfo {author}
  {\bibfnamefont {M.~R.}\ \bibnamefont {Tarbutt}},\ and\ \bibinfo {author}
  {\bibfnamefont {E.~A.}\ \bibnamefont {Hinds}},\ }\bibfield  {title} {\bibinfo
  {title} {Improved measurement of the shape of the electron},\ }\href
  {https://doi.org/10.1038/nature10104} {\bibfield  {journal} {\bibinfo
  {journal} {Nature}\ }\textbf {\bibinfo {volume} {473}},\ \bibinfo {pages}
  {493} (\bibinfo {year} {2011})}\BibitemShut {NoStop}%
\bibitem [{\citenamefont {Kozlov}\ and\ \citenamefont
  {DeMille}(2002)}]{kozlov_enhancement_2002}%
  \BibitemOpen
  \bibfield  {author} {\bibinfo {author} {\bibfnamefont {M.~G.}\ \bibnamefont
  {Kozlov}}\ and\ \bibinfo {author} {\bibfnamefont {D.}~\bibnamefont
  {DeMille}},\ }\bibfield  {title} {\bibinfo {title} {Enhancement of the
  {Electric} {Dipole} {Moment} of the {Electron} in {PbO}},\ }\href
  {https://doi.org/10.1103/PhysRevLett.89.133001} {\bibfield  {journal}
  {\bibinfo  {journal} {Phys. Rev. Lett.}\ }\textbf {\bibinfo {volume} {89}},\
  \bibinfo {pages} {133001} (\bibinfo {year} {2002})}\BibitemShut {NoStop}%
\bibitem [{\citenamefont {Baklanov}\ \emph {et~al.}(2010)\citenamefont
  {Baklanov}, \citenamefont {Petrov}, \citenamefont {Titov},\ and\
  \citenamefont {Kozlov}}]{baklanov_progress_2010}%
  \BibitemOpen
  \bibfield  {author} {\bibinfo {author} {\bibfnamefont {K.~I.}\ \bibnamefont
  {Baklanov}}, \bibinfo {author} {\bibfnamefont {A.~N.}\ \bibnamefont
  {Petrov}}, \bibinfo {author} {\bibfnamefont {A.~V.}\ \bibnamefont {Titov}},\
  and\ \bibinfo {author} {\bibfnamefont {M.~G.}\ \bibnamefont {Kozlov}},\
  }\bibfield  {title} {\bibinfo {title} {Progress toward the electron
  electric-dipole-moment search: {Theoretical} study of the {PbF} molecule},\
  }\href {https://doi.org/10.1103/PhysRevA.82.060501} {\bibfield  {journal}
  {\bibinfo  {journal} {Phys. Rev. A}\ }\textbf {\bibinfo {volume} {82}},\
  \bibinfo {pages} {060501(R)} (\bibinfo {year} {2010})}\BibitemShut {NoStop}%
\bibitem [{\citenamefont {Skripnikov}\ \emph {et~al.}(2014)\citenamefont
  {Skripnikov}, \citenamefont {Kudashov}, \citenamefont {Petrov},\ and\
  \citenamefont {Titov}}]{skripnikov_search_2014}%
  \BibitemOpen
  \bibfield  {author} {\bibinfo {author} {\bibfnamefont {L.~V.}\ \bibnamefont
  {Skripnikov}}, \bibinfo {author} {\bibfnamefont {A.~D.}\ \bibnamefont
  {Kudashov}}, \bibinfo {author} {\bibfnamefont {A.~N.}\ \bibnamefont
  {Petrov}},\ and\ \bibinfo {author} {\bibfnamefont {A.~V.}\ \bibnamefont
  {Titov}},\ }\bibfield  {title} {\bibinfo {title} {Search for parity- and
  time-and-parity--violation effects in lead monofluoride ({PbF}): {Ab} initio
  molecular study},\ }\href {https://doi.org/10.1103/PhysRevA.90.064501}
  {\bibfield  {journal} {\bibinfo  {journal} {Phys. Rev. A}\ }\textbf {\bibinfo
  {volume} {90}},\ \bibinfo {pages} {064501} (\bibinfo {year}
  {2014})}\BibitemShut {NoStop}%
\bibitem [{\citenamefont {Sasmal}\ \emph {et~al.}(2015)\citenamefont {Sasmal},
  \citenamefont {Pathak}, \citenamefont {Nayak}, \citenamefont {Vaval},\ and\
  \citenamefont {Pal}}]{sasmal_calculation_2015}%
  \BibitemOpen
  \bibfield  {author} {\bibinfo {author} {\bibfnamefont {S.}~\bibnamefont
  {Sasmal}}, \bibinfo {author} {\bibfnamefont {H.}~\bibnamefont {Pathak}},
  \bibinfo {author} {\bibfnamefont {M.~K.}\ \bibnamefont {Nayak}}, \bibinfo
  {author} {\bibfnamefont {N.}~\bibnamefont {Vaval}},\ and\ \bibinfo {author}
  {\bibfnamefont {S.}~\bibnamefont {Pal}},\ }\bibfield  {title} {\bibinfo
  {title} {Calculation of {P},{T}-odd interaction constant of {PbF} using
  {Z}-vector method in the relativistic coupled-cluster framework},\ }\href
  {https://doi.org/10.1063/1.4929591} {\bibfield  {journal} {\bibinfo
  {journal} {J. Chem. Phys.}\ }\textbf {\bibinfo {volume} {143}},\ \bibinfo
  {pages} {084119} (\bibinfo {year} {2015})}\BibitemShut {NoStop}%
\bibitem [{\citenamefont {Stuhl}\ \emph {et~al.}(2008)\citenamefont {Stuhl},
  \citenamefont {Sawyer}, \citenamefont {Wang},\ and\ \citenamefont
  {Ye}}]{stuhl2008magnetooptical}%
  \BibitemOpen
  \bibfield  {author} {\bibinfo {author} {\bibfnamefont {B.~K.}\ \bibnamefont
  {Stuhl}}, \bibinfo {author} {\bibfnamefont {B.~C.}\ \bibnamefont {Sawyer}},
  \bibinfo {author} {\bibfnamefont {D.}~\bibnamefont {Wang}},\ and\ \bibinfo
  {author} {\bibfnamefont {J.}~\bibnamefont {Ye}},\ }\bibfield  {title}
  {\bibinfo {title} {Magneto-optical trap for polar molecules},\ }\href
  {https://doi.org/10.1103/PhysRevLett.101.243002} {\bibfield  {journal}
  {\bibinfo  {journal} {Phys. Rev. Lett.}\ }\textbf {\bibinfo {volume} {101}},\
  \bibinfo {pages} {243002} (\bibinfo {year} {2008})},\ \bibinfo {note} {arXiv:
  0808.2171}\BibitemShut {NoStop}%
\bibitem [{\citenamefont {Hamamda}\ \emph {et~al.}(2015)\citenamefont
  {Hamamda}, \citenamefont {Pillet}, \citenamefont {Lignier},\ and\
  \citenamefont {Comparat}}]{hamamda2015rovibrational}%
  \BibitemOpen
  \bibfield  {author} {\bibinfo {author} {\bibfnamefont {M.}~\bibnamefont
  {Hamamda}}, \bibinfo {author} {\bibfnamefont {P.}~\bibnamefont {Pillet}},
  \bibinfo {author} {\bibfnamefont {H.}~\bibnamefont {Lignier}},\ and\ \bibinfo
  {author} {\bibfnamefont {D.}~\bibnamefont {Comparat}},\ }\bibfield  {title}
  {\bibinfo {title} {Ro-vibrational cooling of molecules and prospects},\
  }\href {https://doi.org/10.1088/0953-4075/48/18/182001} {\bibfield  {journal}
  {\bibinfo  {journal} {J. Phys. B}\ }\textbf {\bibinfo {volume} {48}},\
  \bibinfo {pages} {22796} (\bibinfo {year} {2015})}\BibitemShut {NoStop}%
\bibitem [{\citenamefont {Isaev}\ and\ \citenamefont
  {Berger}(2016)}]{isaev2016polyatomic}%
  \BibitemOpen
  \bibfield  {author} {\bibinfo {author} {\bibfnamefont {T.~A.}\ \bibnamefont
  {Isaev}}\ and\ \bibinfo {author} {\bibfnamefont {R.}~\bibnamefont {Berger}},\
  }\bibfield  {title} {\bibinfo {title} {Polyatomic candidates for cooling of
  molecules with lasers from simple theoretical concepts},\ }\href
  {https://doi.org/10.1103/PhysRevLett.116.063006} {\bibfield  {journal}
  {\bibinfo  {journal} {Phys. Rev. Lett.}\ }\textbf {\bibinfo {volume} {116}},\
  \bibinfo {pages} {063006} (\bibinfo {year} {2016})}\BibitemShut {NoStop}%
\bibitem [{\citenamefont {Kozyryev}\ \emph {et~al.}(2016)\citenamefont
  {Kozyryev}, \citenamefont {Baum}, \citenamefont {Matsuda},\ and\
  \citenamefont {Doyle}}]{kozyryev2016proposal}%
  \BibitemOpen
  \bibfield  {author} {\bibinfo {author} {\bibfnamefont {I.}~\bibnamefont
  {Kozyryev}}, \bibinfo {author} {\bibfnamefont {L.}~\bibnamefont {Baum}},
  \bibinfo {author} {\bibfnamefont {K.}~\bibnamefont {Matsuda}},\ and\ \bibinfo
  {author} {\bibfnamefont {J.~M.}\ \bibnamefont {Doyle}},\ }\bibfield  {title}
  {\bibinfo {title} {Proposal for {Laser} {Cooling} of {Complex} {Polyatomic}
  {Molecules}},\ }\href {https://doi.org/10.1002/cphc.201601051} {\bibfield
  {journal} {\bibinfo  {journal} {ChemPhysChem}\ }\textbf {\bibinfo {volume}
  {17}},\ \bibinfo {pages} {3641} (\bibinfo {year} {2016})}\BibitemShut
  {NoStop}%
\bibitem [{\citenamefont {Augenbraun}\ \emph
  {et~al.}(2020{\natexlab{b}})\citenamefont {Augenbraun}, \citenamefont
  {Doyle}, \citenamefont {Zelevinsky},\ and\ \citenamefont
  {Kozyryev}}]{augenbraun2020molecular}%
  \BibitemOpen
  \bibfield  {author} {\bibinfo {author} {\bibfnamefont {B.~L.}\ \bibnamefont
  {Augenbraun}}, \bibinfo {author} {\bibfnamefont {J.~M.}\ \bibnamefont
  {Doyle}}, \bibinfo {author} {\bibfnamefont {T.}~\bibnamefont {Zelevinsky}},\
  and\ \bibinfo {author} {\bibfnamefont {I.}~\bibnamefont {Kozyryev}},\
  }\bibfield  {title} {\bibinfo {title} {Molecular {Asymmetry} and {Optical}
  {Cycling}: {Laser} {Cooling} {Asymmetric} {Top} {Molecules}},\ }\href
  {https://doi.org/10.1103/PhysRevX.10.031022} {\bibfield  {journal} {\bibinfo
  {journal} {Phys. Rev. X}\ }\textbf {\bibinfo {volume} {10}},\ \bibinfo
  {pages} {031022} (\bibinfo {year} {2020}{\natexlab{b}})}\BibitemShut
  {NoStop}%
\bibitem [{\citenamefont {Vilas}\ \emph {et~al.}(2022)\citenamefont {Vilas},
  \citenamefont {Hallas}, \citenamefont {Anderegg}, \citenamefont {Robichaud},
  \citenamefont {Winnicki}, \citenamefont {Mitra},\ and\ \citenamefont
  {Doyle}}]{vilas_magneto-optical_2022}%
  \BibitemOpen
  \bibfield  {author} {\bibinfo {author} {\bibfnamefont {N.~B.}\ \bibnamefont
  {Vilas}}, \bibinfo {author} {\bibfnamefont {C.}~\bibnamefont {Hallas}},
  \bibinfo {author} {\bibfnamefont {L.}~\bibnamefont {Anderegg}}, \bibinfo
  {author} {\bibfnamefont {P.}~\bibnamefont {Robichaud}}, \bibinfo {author}
  {\bibfnamefont {A.}~\bibnamefont {Winnicki}}, \bibinfo {author}
  {\bibfnamefont {D.}~\bibnamefont {Mitra}},\ and\ \bibinfo {author}
  {\bibfnamefont {J.~M.}\ \bibnamefont {Doyle}},\ }\bibfield  {title} {\bibinfo
  {title} {Magneto-optical trapping and sub-{Doppler} cooling of a polyatomic
  molecule},\ }\href {https://doi.org/10.1038/s41586-022-04620-5} {\bibfield
  {journal} {\bibinfo  {journal} {Nature}\ }\textbf {\bibinfo {volume} {606}},\
  \bibinfo {pages} {70} (\bibinfo {year} {2022})}\BibitemShut {NoStop}%
\bibitem [{\citenamefont {Zeng}\ \emph {et~al.}(2024)\citenamefont {Zeng},
  \citenamefont {Deng}, \citenamefont {Yang},\ and\ \citenamefont
  {Yan}}]{zeng_three-dimensional_2024}%
  \BibitemOpen
  \bibfield  {author} {\bibinfo {author} {\bibfnamefont {Z.}~\bibnamefont
  {Zeng}}, \bibinfo {author} {\bibfnamefont {S.}~\bibnamefont {Deng}}, \bibinfo
  {author} {\bibfnamefont {S.}~\bibnamefont {Yang}},\ and\ \bibinfo {author}
  {\bibfnamefont {B.}~\bibnamefont {Yan}},\ }\bibfield  {title} {\bibinfo
  {title} {Three-dimensional {Magneto}-optical {Trapping} of {Barium}
  {Monofluoride}},\ }\href {http://arxiv.org/abs/2405.17883} {\bibfield
  {journal} {\bibinfo  {journal} {arXiv:2405.17883}\ } (\bibinfo {year}
  {2024})}\BibitemShut {NoStop}%
\bibitem [{\citenamefont {Schnaubelt}\ \emph {et~al.}(2021)\citenamefont
  {Schnaubelt}, \citenamefont {Shaw},\ and\ \citenamefont
  {McCarron}}]{schnaubelt2021cold}%
  \BibitemOpen
  \bibfield  {author} {\bibinfo {author} {\bibfnamefont {J.~C.}\ \bibnamefont
  {Schnaubelt}}, \bibinfo {author} {\bibfnamefont {J.~C.}\ \bibnamefont
  {Shaw}},\ and\ \bibinfo {author} {\bibfnamefont {D.~J.}\ \bibnamefont
  {McCarron}},\ }\bibfield  {title} {\bibinfo {title} {Cold {CH} radicals for
  laser cooling and trapping},\ }\href {https://arxiv.org/abs/2109.03953}
  {\bibfield  {journal} {\bibinfo  {journal} {arXiv:2109.03953}\ } (\bibinfo
  {year} {2021})}\BibitemShut {NoStop}%
\bibitem [{\citenamefont {Zeng}\ \emph {et~al.}(2023)\citenamefont {Zeng},
  \citenamefont {Jadbabaie}, \citenamefont {Patel}, \citenamefont {Yu},
  \citenamefont {Steimle},\ and\ \citenamefont {Hutzler}}]{zeng_optical_2023}%
  \BibitemOpen
  \bibfield  {author} {\bibinfo {author} {\bibfnamefont {Y.}~\bibnamefont
  {Zeng}}, \bibinfo {author} {\bibfnamefont {A.}~\bibnamefont {Jadbabaie}},
  \bibinfo {author} {\bibfnamefont {A.~N.}\ \bibnamefont {Patel}}, \bibinfo
  {author} {\bibfnamefont {P.}~\bibnamefont {Yu}}, \bibinfo {author}
  {\bibfnamefont {T.~C.}\ \bibnamefont {Steimle}},\ and\ \bibinfo {author}
  {\bibfnamefont {N.~R.}\ \bibnamefont {Hutzler}},\ }\bibfield  {title}
  {\bibinfo {title} {Optical cycling in polyatomic molecules with complex
  hyperfine structure},\ }\href {https://doi.org/10.1103/PhysRevA.108.012813}
  {\bibfield  {journal} {\bibinfo  {journal} {Phys. Rev. A}\ }\textbf {\bibinfo
  {volume} {108}},\ \bibinfo {pages} {012813} (\bibinfo {year}
  {2023})}\BibitemShut {NoStop}%
\bibitem [{\citenamefont {Khriplovich}\ and\ \citenamefont
  {Lamoreaux}(1997)}]{khriplovich_cp_1997}%
  \BibitemOpen
  \bibfield  {author} {\bibinfo {author} {\bibfnamefont {I.~B.}\ \bibnamefont
  {Khriplovich}}\ and\ \bibinfo {author} {\bibfnamefont {S.~K.}\ \bibnamefont
  {Lamoreaux}},\ }\href {https://doi.org/10.1007/978-3-642-60838-4} {\emph
  {\bibinfo {title} {{CP} Violation Without Strangeness}}}\ (\bibinfo
  {publisher} {Springer Berlin Heidelberg},\ \bibinfo {year}
  {1997})\BibitemShut {NoStop}%
\bibitem [{\citenamefont {Wu}\ \emph {et~al.}(2020)\citenamefont {Wu},
  \citenamefont {Han}, \citenamefont {Chow}, \citenamefont {Ang}, \citenamefont
  {Meisenhelder}, \citenamefont {Panda}, \citenamefont {West}, \citenamefont
  {Gabrielse}, \citenamefont {Doyle},\ and\ \citenamefont
  {DeMille}}]{wu_metastable_2020}%
  \BibitemOpen
  \bibfield  {author} {\bibinfo {author} {\bibfnamefont {X.}~\bibnamefont
  {Wu}}, \bibinfo {author} {\bibfnamefont {Z.}~\bibnamefont {Han}}, \bibinfo
  {author} {\bibfnamefont {J.}~\bibnamefont {Chow}}, \bibinfo {author}
  {\bibfnamefont {D.~G.}\ \bibnamefont {Ang}}, \bibinfo {author} {\bibfnamefont
  {C.}~\bibnamefont {Meisenhelder}}, \bibinfo {author} {\bibfnamefont {C.~D.}\
  \bibnamefont {Panda}}, \bibinfo {author} {\bibfnamefont {E.~P.}\ \bibnamefont
  {West}}, \bibinfo {author} {\bibfnamefont {G.}~\bibnamefont {Gabrielse}},
  \bibinfo {author} {\bibfnamefont {J.~M.}\ \bibnamefont {Doyle}},\ and\
  \bibinfo {author} {\bibfnamefont {D.}~\bibnamefont {DeMille}},\ }\bibfield
  {title} {\bibinfo {title} {The metastable
  {Q}\${\textasciicircum}3{\textbackslash}{Delta}\_2\$ state of {ThO}: a new
  resource for the {ACME} electron {EDM} search},\ }\href
  {https://doi.org/10.1088/1367-2630/ab6a3a} {\bibfield  {journal} {\bibinfo
  {journal} {New J. Phys.}\ }\textbf {\bibinfo {volume} {22}},\ \bibinfo
  {pages} {023013} (\bibinfo {year} {2020})}\BibitemShut {NoStop}%
\bibitem [{\citenamefont {Verma}\ \emph {et~al.}(2020)\citenamefont {Verma},
  \citenamefont {Jayich},\ and\ \citenamefont {Vutha}}]{verma_electron_2020}%
  \BibitemOpen
  \bibfield  {author} {\bibinfo {author} {\bibfnamefont {M.}~\bibnamefont
  {Verma}}, \bibinfo {author} {\bibfnamefont {A.~M.}\ \bibnamefont {Jayich}},\
  and\ \bibinfo {author} {\bibfnamefont {A.~C.}\ \bibnamefont {Vutha}},\
  }\bibfield  {title} {\bibinfo {title} {Electron {Electric} {Dipole} {Moment}
  {Searches} {Using} {Clock} {Transitions} in {Ultracold} {Molecules}},\ }\href
  {https://doi.org/10.1103/PhysRevLett.125.153201} {\bibfield  {journal}
  {\bibinfo  {journal} {Phys. Rev. Lett.}\ }\textbf {\bibinfo {volume} {125}},\
  \bibinfo {pages} {153201} (\bibinfo {year} {2020})}\BibitemShut {NoStop}%
\bibitem [{\citenamefont {Augenbraun}\ \emph {et~al.}(2021)\citenamefont
  {Augenbraun}, \citenamefont {Frenett}, \citenamefont {Sawaoka}, \citenamefont
  {Hallas}, \citenamefont {Vilas}, \citenamefont {Nasir}, \citenamefont
  {Lasner},\ and\ \citenamefont {Doyle}}]{augenbraun2021zeemansisyphus}%
  \BibitemOpen
  \bibfield  {author} {\bibinfo {author} {\bibfnamefont {B.~L.}\ \bibnamefont
  {Augenbraun}}, \bibinfo {author} {\bibfnamefont {A.}~\bibnamefont {Frenett}},
  \bibinfo {author} {\bibfnamefont {H.}~\bibnamefont {Sawaoka}}, \bibinfo
  {author} {\bibfnamefont {C.}~\bibnamefont {Hallas}}, \bibinfo {author}
  {\bibfnamefont {N.~B.}\ \bibnamefont {Vilas}}, \bibinfo {author}
  {\bibfnamefont {A.}~\bibnamefont {Nasir}}, \bibinfo {author} {\bibfnamefont
  {Z.~D.}\ \bibnamefont {Lasner}},\ and\ \bibinfo {author} {\bibfnamefont
  {J.~M.}\ \bibnamefont {Doyle}},\ }\bibfield  {title} {\bibinfo {title}
  {Zeeman-{Sisyphus} {Deceleration} of {Molecular} {Beams}},\ }\href
  {https://doi.org/10.1103/PhysRevLett.127.263002} {\bibfield  {journal}
  {\bibinfo  {journal} {Phys. Rev. Lett.}\ }\textbf {\bibinfo {volume} {127}},\
  \bibinfo {pages} {263002} (\bibinfo {year} {2021})}\BibitemShut {NoStop}%
\bibitem [{\citenamefont {Fitch}\ and\ \citenamefont
  {Tarbutt}(2016)}]{fitch2016principles}%
  \BibitemOpen
  \bibfield  {author} {\bibinfo {author} {\bibfnamefont {N.~J.}\ \bibnamefont
  {Fitch}}\ and\ \bibinfo {author} {\bibfnamefont {M.~R.}\ \bibnamefont
  {Tarbutt}},\ }\bibfield  {title} {\bibinfo {title} {Principles and {Design}
  of a {Zeeman}–{Sisyphus} {Decelerator} for {Molecular} {Beams}},\ }\href
  {https://doi.org/10.1002/cphc.201600656} {\bibfield  {journal} {\bibinfo
  {journal} {ChemPhysChem}\ }\textbf {\bibinfo {volume} {17}},\ \bibinfo
  {pages} {3609} (\bibinfo {year} {2016})}\BibitemShut {NoStop}%
\bibitem [{\citenamefont {Comparat}(2014)}]{comparat2014molecular}%
  \BibitemOpen
  \bibfield  {author} {\bibinfo {author} {\bibfnamefont {D.}~\bibnamefont
  {Comparat}},\ }\bibfield  {title} {\bibinfo {title} {Molecular cooling via
  {Sisyphus} processes},\ }\href {https://doi.org/10.1103/PhysRevA.89.043410}
  {\bibfield  {journal} {\bibinfo  {journal} {Phys. Rev. A}\ }\textbf {\bibinfo
  {volume} {89}},\ \bibinfo {pages} {043410} (\bibinfo {year}
  {2014})}\BibitemShut {NoStop}%
\bibitem [{\citenamefont {Zeppenfeld}\ \emph {et~al.}(2012)\citenamefont
  {Zeppenfeld}, \citenamefont {Englert}, \citenamefont {Glöckner},
  \citenamefont {Prehn}, \citenamefont {Mielenz}, \citenamefont {Sommer},
  \citenamefont {Van~Buuren}, \citenamefont {Motsch},\ and\ \citenamefont
  {Rempe}}]{zeppenfeld2012sisyphus}%
  \BibitemOpen
  \bibfield  {author} {\bibinfo {author} {\bibfnamefont {M.}~\bibnamefont
  {Zeppenfeld}}, \bibinfo {author} {\bibfnamefont {B.~G.~U.}\ \bibnamefont
  {Englert}}, \bibinfo {author} {\bibfnamefont {R.}~\bibnamefont {Glöckner}},
  \bibinfo {author} {\bibfnamefont {A.}~\bibnamefont {Prehn}}, \bibinfo
  {author} {\bibfnamefont {M.}~\bibnamefont {Mielenz}}, \bibinfo {author}
  {\bibfnamefont {C.}~\bibnamefont {Sommer}}, \bibinfo {author} {\bibfnamefont
  {L.~D.}\ \bibnamefont {Van~Buuren}}, \bibinfo {author} {\bibfnamefont
  {M.}~\bibnamefont {Motsch}},\ and\ \bibinfo {author} {\bibfnamefont
  {G.}~\bibnamefont {Rempe}},\ }\bibfield  {title} {\bibinfo {title} {Sisyphus
  cooling of electrically trapped polyatomic molecules},\ }\href
  {https://doi.org/10.1038/nature11595} {\bibfield  {journal} {\bibinfo
  {journal} {Nature}\ }\textbf {\bibinfo {volume} {491}},\ \bibinfo {pages}
  {570} (\bibinfo {year} {2012})}\BibitemShut {NoStop}%
\end{thebibliography}%

\end{document}